\documentclass[a4paper,fleqn,usenatbib]{mnras}

\usepackage{newtxtext,newtxmath}

\usepackage[T1]{fontenc}
\usepackage{ae,aecompl}
\usepackage{relsize} 
\usepackage{color,soul}
\usepackage{subfig}

\usepackage{graphicx}
\usepackage{amsmath}
\usepackage{amssymb}
\usepackage{bm}
\usepackage{footnote}
\usepackage{bigints}
\usepackage{etoolbox}
\makeatletter
\AtBeginEnvironment{thebibliography}{%
   \clubpenalty10000
   \@clubpenalty \clubpenalty
   \widowpenalty10000}
\makeatother
\defcitealias{Eis18}{E18}

\title[External photoevaporation of PPDs in the ONC]{\vspace{-2mm}A solution to the proplyd lifetime problem\vspace{-3mm}}

\author[A.~J.~Winter \textit{et al.}]{Andrew~J.~Winter,$^{1,2}$\thanks{a.j.winter@le.ac.uk}	
 Cathie~J.~Clarke$^{2}$, Giovanni~P.~Rosotti$^{3}$, Alvaro Hacar$^{3}$, \newauthor Richard Alexander$^{1}$\\
$^{1}$Department of Physics and Astronomy, University of Leicester, Leicester, LE1 7RH, UK\\
$^{2}$Institute of Astronomy, University of Cambridge, Madingley Road, Cambridge CB3 0HA, UK \\
$^{3}$Leiden Observatory, Leiden University, P.O. Box 9513, 2300-RA Leiden, The Netherlands
\vspace{-3mm}
}

\date{Accepted 2019 {September} 9. Received 2019 {September} 6; in original form 2019 July 26}\vspace{-2mm}

\pubyear{2019}

\begin{document}
\label{firstpage}
\pagerange{\pageref{firstpage}--\pageref{lastpage}}
\maketitle

\begin{abstract}
Protoplanetary discs (PPDs) in the Orion Nebula Cluster (ONC) are irradiated by UV fields from the massive star $\theta^1$C. This drives thermal winds, inducing mass loss rates of up to $\dot{M}_\mathrm{wind}\sim 10^{-7}\,M_\odot$~yr$^{-1}$ in the `proplyds' (ionised PPDs) close to the centre. For the mean age of the ONC and reasonable initial PPD masses, such mass loss rates imply that discs should have been dispersed. However, $\sim 80\%$ of stars still exhibit a NIR excess, suggesting that significant circumstellar mass remains. This `proplyd lifetime problem' has persisted since the discovery of photoevaporating discs in the core of the ONC by \citet{Ode94}. In this work, we demonstrate how an extended period of star formation can solve this problem. Coupling $N$-body calculations and a viscous disc evolution model, we obtain high disc fractions at the present day. This is partly due to the migration of older stars outwards, and younger stars inwards such that the most strongly irradiated PPDs are also the youngest. We show how the disc mass distribution can be used to test the recent claims in the literature for multiple stellar populations in the ONC. Our model also explains the recent finding that host mass and PPD mass are only weakly correlated, in contrast with other regions of similar age. We conclude that the status of the ONC as the archetype for understanding the influence of environment on planet formation is undeserved; the complex star formation history (involving star formation episodes within $\sim 0.8$~Myr of the present day) results in confusing signatures in the PPD population. 
\end{abstract}

\begin{keywords} 
stars: formation, circumstellar matter, kinematics and dynamics -- open clusters and associations: individual: Orion Nebula Cluster -- protoplanetary discs\vspace{-2mm}
\end{keywords}


\section{Introduction}
Protoplanetary discs (PPDs) are discs of dust and gas around a young star, which is the material available for the formation of planets. The time permitted for planet formation is therefore limited by the lifetime of PPDs. The dispersal timescale for discs in relatively isolated environments is empirically found to be $\sim 3$~Myr \citep{Hai01b,Fed10,Ing12,Rib15}, which approximately coincides with the formation timescale for giant planets by core accretion \citep[see][for a recent review]{Hel14}. For discs which are dispersed more rapidly than this, giant planet formation may be curtailed. Giant planets play an important role in planetary dynamics and the distribution of molecules towards the inner terrestrial planets \citep[e.g][]{Cha01,Bat15, Agn18}. Hence, quantifying the factors that determine PPD dispersal are important in understanding observed exoplanet architectures and chemistry.

A growing body of work suggests that planet formation is strongly dependent on the birth environment of the host star. Stars preferentially form in groups \citep{Lad03}, and in sufficiently dense environments the evolution of a PPD can be significantly influenced by neighbours \citep{dJO12}. Close star-disc encounters are one such environmental influence on PPDs that can result in enhanced accretion and hasten disc depletion \citep{Cla93,Ost94,Pfa05,Olc06,Win18,Bat18,Cue19}. However, the stellar number densities required for tidal truncation are high, and in practice few observed regions satisfy this condition \citep{Win18b, Win19b}. The influence of tidal truncation is therefore limited to stellar multiples, either in bound systems \citep{Dai15, Kur18} or during the decay of higher order multiplicity \citep{Win18c}. Since stellar multiplicity does not appear to be strongly dependent on environment \mbox{\citep[see][for a review]{Duc13}}, this suggests that encounters are not an environmental influence, but may set disc initial conditions during the early phases of cluster evolution \citep{Bat18}. Discs can also be externally depleted via thermal winds driven by far ultraviolet (FUV) and extreme ultraviolet (EUV) photons from neighbouring massive stars \citep{Joh98,Sto99, Ada04, Fac16, Haw18b, Haw19}. This process of external photoevaporation dominates over dynamical encounters in observed environments, and can deplete PPDs rapidly for many stars that are born in massive and dense clustered environments \citep{Sca01,Win18b}. Many stars in the solar neighbourhood are born in regions where UV fields are sufficient to significantly shorten disc lifetimes \citep[][]{Fat08, Win18b}, and the fraction of stars born in such environments may be much greater outside of this region, dependent on galactic environment \citep{Win19b}.

The first observational evidence of the external photoevaporation of PPDs were the `proplyds' in the Orion Nebula Cluster \citep[ONC --][]{Ode94}. These comet-like structures are PPDs that are exposed to strong FUV and EUV flux ($F_\mathrm{FUV} \gtrsim 10^4 \, G_0$\footnote{$1\,G_0\equiv 1.6\times 10^{-3}$~erg~cm$^{-2}$~s$^{-1}$ is the \citet{Hab68} unit, the mean FUV flux in the solar neighbourhood.}) originating from the O star $\theta^1$C Ori. \citet{Joh98} explained the morphology of proplyds as the illuminated ionisation front surrounding photoevaporating discs, which experience significant mass loss due to thermal winds. The mass loss rates for proplyds inferred from theory \citep{Joh98, Sto99} and observations \citep{Hen98, Hen99, Hen02} are $\sim 10^{-8}$--$10^{-6}\,M_\odot$~yr$^{-1}$ for PPDs close ($\lesssim 0.3$~pc) to $\theta^1$C. Such rapid mass loss suggests that either the discs were initially extremely massive ($\gtrsim\, 1\, M_\odot$) or $\theta^1$C Ori is very young ($\lesssim 0.1$~Myr). The former explanation is physically implausible since discs with masses greater than the host star are dynamically unstable and short-lived. Indeed, typical disc masses in young PPD samples are $\lesssim 0.1\, M_\odot$ \citep[e.g.][]{Taz17, Ans17}, based on standard assumptions on dust-to-gas ratio and mm opacity. The latter explanation, while possible, would mean a much younger age for $\theta^1$C than the rest of the stellar population; for example, \citet{DRi10} find the stellar age distribution in the ONC peak at $\sim 2$--$3$~Myr. Continuum observations at wavelengths where dust emission dominates over the free-free indicate that discs in the ONC are low mass, with the majority having a total mass $M_\mathrm{disc}\lesssim 0.01\, M_\odot$ \citep[assuming a fiducial dust-to-gas ratio of $10^{-2}$ --][]{Man14, Eis18}. Such a low mass would imply that we are observing the ONC at a special time, over the short $\sim 0.1$~Myr period where discs are strongly irradiated and low mass but are still present around the majority of stars. This uncomfortable coincidence is known as the `proplyd lifetime problem'.

The focus of this work is to address the proplyd lifetime problem from a modelling perspective using recent developments in both the theory of photoevaporating discs and observations of the star and disc population in the ONC. The most similar study to this work is that of \citet{Sca01}, who produced an $N$-body model of the ONC coupled with theoretical photoevaporative mass loss rates. The aim was to test the idea put forward by \mbox{\citet{Sto99}} that radial orbits of stars could result in shorter periods of exposure to strong FUV flux close to $\theta^1$C, thus increasing the PPD dispersal timescale. The models demonstrated that such dynamical orbits alone are insufficient to produce significantly extended disc lifetimes. However, since that study a number of developments mean that the problem is due to be revisited. Firstly, thermodynamic calculations of conditions in photodissociation regions (PDRs) have been coupled with self-consistent equations for the thermally driven disc wind \citep{Ada04, Fac16, Haw18b, Haw19}. In particular, the recent \textsc{Fried} grid \citep{Haw18b} can be interpolated across FUV flux, disc outer radius, and disc mass to calculate an instantaneous mass loss rate for a given PPD. Secondly, the recent sub-mm survey of disc masses and radii by \citet[][hereafter \citetalias{Eis18}]{Eis18} offers further observational constraints on a successful model of PPD evolution, and the host-mass dependent disc properties permit a test of theoretical predictions. Thirdly, \citet[][see also \citealt{Jer19}]{Bec17} recently found evidence of three stellar populations with distinct ages in the ONC. This has multiple consequences for models of a PPD population, most obviously that a subset of discs has evolved for a shorter period than the oldest stars in the region. Additionally, since stars are formed from gas, during the evolution of the cluster prior to the most recent formation event there may have been considerable intra-cluster extinction of UV photons. Understanding how such a scenario infuences the disc population would also represent a test of the hypothesis that discrete epochs of star formation occured in the ONC.

In the following work, we first review the properties of the ONC in Section~\ref{sec:ONCprops}. In Section \ref{sec:nummethod} we summarise our numerical method and modelling approachs. We present our results in Section~\ref{sec:resanddisc}, where for each aspect of our model we also draw comparisons to recent works and discuss the consequences of our findings. The composite solution to the proplyd lifetime problem and caveats to our model are summarised in Section~\ref{sec:PLP_sol}. Our main conclusions are drawn in Section~\ref{sec:concs}.

 \section{Properties of the ONC}
 \label{sec:ONCprops}
 
In the plane of the sky the ONC lies on the Orion A molecular cloud, a star forming region with a filamentary morphology and a mass of $\sim 5\times 10^4\, M_\odot$ \citep{Lom14}. \citet{Ment07} used \textit{Very Long Baseline Array} (VLBA) measurements to calculate a distance of  $414\pm 7$~pc \citep[see also][]{Hir07, Kim08}. Since the ONC exhibits relatively low extinction \citep[$A_V \sim 1.5^m$ --][]{Joh67, ODe00, DRi10} in the direction of its members, it must lie physically in front of the Orion Molecular Cloud (OMC), which can reach up to $A_V \sim 50^m$ \citep{Ber96, Sca11, Lom11}. In this section we review some pertinent empirically constrained properties of the stellar and disc population.


\subsection{Stellar mass and density profile}

The distribution of stars in the ONC appears slightly elliptical on the plane of the sky, extended along the north/south direction that traces the orientation of the Orion A molecular filament. However, for convenience most studies seeking to quantify the mass density profile assume spherical symmetry. \mbox{\citet{Hil98}} applied a King model to the observed population and found a central density of $\sim 2 \times 10^4$~stars~pc$^{-3}$ and a core radius of $\sim 0.2$~pc \citep[see also][]{DaR14}. The total stellar mass is $\sim 4000\, M_\odot$ inside $3$~pc (although the definition of the ONC in terms of radial extent varies between studies).

There is evidence of mass segregation in the ONC, with the most massive stars found in the central regions \mbox{\citep{Hil97}}. Since the half-mass relaxation time \mbox{\citep{Spi71}} for the ONC is $\tau_{\mathrm{rh}} \gtrsim 5$~Myr, it is unlikely that the stellar population would segregate in mass by its current age without an epoch of violent relaxation \mbox{\citep{Bon98, All09}}. This suggests that either the ONC was primordially mass segregated, or a period of collapse occurred early during its evolution. In support of the latter hypothesis, \mbox{\citet{Kuz18}} find that cold collapse can reproduce kinematic signatures in the gas and stellar population similar to those around Orion A found by \mbox{\citet{DaR17}}.  

 \subsection{Stellar kinematics}
 
The stars in the ONC have been the subject of a number of proper motion studies \citep{Par54,Str58,Jon88, vAl88, Dzi17, Kuh19}. Most recently, \citet{Kim19} used long baseline \textit{Hubble Space Telescope} (HST) observations and high resolution near-IR data from the Keck II telescope to achieve high precision proper motion measurements for $701$ cluster members. They find that the velocity dispersion in RA and Dec are $\sigma_{\alpha}=1.57\pm 0.04$~km/s and $\sigma_{\delta}=2.12\pm
0.06$~km/s respectively. The greater dispersion along in the north/south direction \citep[see also][]{Jon88, Kuh19}, aligned with the Orion A molecular cloud, may be the result of the filaments' contribution to the gravitational potential.


Whether or not the ONC is expanding or contracting has historically been the subject of debate \citep[see e.g.][]{Mue08}. Virial equilibrium in the ONC requires a one-dimensional velocity dispersion of $\sigma_v \approx 1.6$~km/s \citep{DaR14}, comparable to the observed value. Recently, \citet{Kim19} found no evidence of expansion by considering the mean radial component of the stellar proper motions \citep[and discusses the contrary claim by][]{Kuh19}. Overall, the evidence suggests that the ONC is close to virial equilibrium at the present time.


 \subsection{Stellar ages}
 
 Non-uniform extinction in the direction of the ONC \citep[e.g.][]{DRi10}, as well as foreground stars which may be spatially distinct and older \citep{Alv12, Bou14}, complicate the determination of the stellar ages. The early study by \citet{Hil97} found a mean age of $\sim 1$~Myr, with a spread of $\sim 2$~Myr, with the youngest stars occupying the central regions (which coincides with the region close to $\theta^1$C, where proplyds are found). Subsequently, \citet{Pal99} find stellar ages peak at $\sim 2$~Myr, with an accelerated phase of star formation in the last $\sim 3$~Myr, but with slower star formation also extending to $\sim 10$~Myr ago. However, \citet{Jef11} showed that the inferred large spread in age is incompatible with the disc fractions (fraction of stars with near infrared excess) as a function of stellar age for reasonable disc lifetimes.
 
 
 Most recently, the \textit{OmegaCAM} survey by \citet[][see also \citealt{Jer19}]{Bec17} found three distinct populations of stars in the distribution of $r-i$ colours, which cannot be attributed to variable extinction or projected distances. The authors claim the finding could either be due to three populations of different age or a previously undiscovered population of tight binaries which are skewed towards an unusually high mass ratio and make up $\sim 50\%$ of the population. To test the former hypothesis, the authors use the ages derived by \citet{DaR16} and find a clear distinction between the populations. They conclude that the most likely explanation is that three discrete periods of star formation occurred in the ONC with $1$-$\sigma$ age ranges $2.51$--$3.28$, $1.55$--$2.29$ and $1.08$--$1.53~$~Myr. \citet[][]{Kro18} also use isochrones by \citet{Bre12} to give a slightly younger age of $0.8$~Myr for the youngest population (while the other two remain largely consistent). The number of stars decreases by a factor $\sim 2$ in each generation chronologically. In agreement with \citet{Hil97}, \citet{Bec17} also found that the youngest stellar population occupies the centre of the ONC \citep[see also][]{Reg11, Jer19}. \citet{Get14} also arrived at the same conclusion using X-ray and near infrared (NIR) photometry.

 \subsection{PPD properties}
 
 \citet{Jef11} reviews a number of previous studies that aimed to quantify the fraction of surviving discs in the ONC. The early investigation by \citet{Hil98b} used the criterion that the $I-K$ magnitude was $>0.3^m$ redder than expected from the spectral type, yielding a disc fraction in the range $55$--$90\%$. The observations covered $\sim 700$~arcmin$^2$ ($\sim 10$~pc$^2$) with some central concentration. \citet{Lad00} presented results from a much more centrally concentrated sample covering $36$~arcmin$^2$ ($0.5$~pc$^2$). Using a reddening criteria in the $J-H$ versus $K-L$ magnitudes, they found a consistent disc fraction, $115$ with and $35$ without ($\sim 77\%$). 
 
Constraints on PPD dust masses and radii are possible from mm/sub-mm observations, since at these wavelengths the dust is optically thin. \citet{Mun95} presented $16$ detections in the ONC at $3.5$~mm, however the free-free emission at this wavelength frequently dominates the detected flux density. Since then, a number of studies have imaged discs in a range of wavelengths such that the dust emission can be distinguished from the free-free component, indicating disc masses in the region are largely $\lesssim 10^{-2}\, M_\odot$ \citep{Bal98, Wil05, Man14, Eis16}. Most recently, \citetalias{Eis18} used the \textit{Atacama Large Millimeter Array} (ALMA) to observe the PPD population in the very central regions (a radius of $\sim 0.17$~pc around $\theta^1$C) at $850$~$\mu$m with the highest sensitivity ($\sim 0.1$~mJy, corresponding to a $4$-$\sigma$ detection limit of $\sim 1\, M_\oplus$ of dust) and resolution ($0''.1$, $\sim 40$~au) to date. Detection fractions are quoted in that study as a fraction of previously identified NIR targets and proplyds. Of the proplyds, $65\%$ exhibit (detected) continuum flux in excess of the free-free component, while this falls to $45\%$ when all members are included. The maximum dust masses are found to be $\sim 80 \, M_\oplus$, while only a weak correlation is found between PPD mass and distance from $\theta^1$C, which is not surprising given that the crossing time in the core is $\ll 1$~Myr (combined with, for example, projection effects). However, \citet{vTer19} have found that discs at much larger ($\sim 1$~pc) separations exhibit greater dust masses by a factor $\sim 5$. In the \citetalias{Eis18} sample, PPD mass is a much shallower function of stellar mass than found in other regions such as Taurus \citep{Andr13}, Chameleon \citep{Pas16}, Upper Sco \citep{Bar16}, Lupus \citep{Ans16} and $\sigma$-Orionis \citep[for a summary see Figure 7 in][]{Ans17}. This is ostensibly a surprising result, particularly as external photoevaporation is expected to preferentially deplete PPDs around low-mass stars \citep{Haw18b, Win19}. Disc radius and host mass are found to be (weakly) positively correlated, which is interpreted to be due to the larger gravitational radius \citep{Hol94}:
\begin{equation}
    R_{g} = \frac{Gm_*}{c_\mathrm{s}^2},
\end{equation} (the radius at which the ionised gas with sound speed $c_\mathrm{s}\approx 10$~km/s is gravitationally unbound) for higher mass stars.

 \subsection{Star formation history hypotheses}

Many studies have aimed to explain the formation history of the ONC. For example, \citet{Hart07} showed that an elliptical rotating sheet with a moderate density gradient can collapse into a structure which resembles Orion A, and could form concentrations of stars that resemble the ONC \citep[see also][]{Kuz18}.  \citet[][see also \citealt{Bou14}]{Alv12} highlight that star formation in the ONC could have been triggered by supernovae in neighbouring regions. Alternatively, \citet{Fuk18} indicate that gas kinematics may support a past collision between molecular clouds.

 However, none of the above scenarios directly explain the recent observations by \citet{Bec17} suggesting discrete epochs of star formation within the ONC. \citet[][see also \citealt{Wan19}]{Kro18} put forward a physical mechanism by which an independently evolving, intermediate mass primordial cluster could self-regulate star formation in such a way as to produce discrete formation events. The authors point to evidence of runaway stars in the ONC \citep{Pov05}, and suggest that previously formed massive stars, which may have suppressed star formation by ionising the gas, have been similarly ejected. Once ejected, some fraction of the gas may have survived, or it may be replenished along the filaments of Orion A \citep[as suggested by the gas kinematics --][]{Hac17}. This is possible only for cluster masses $\sim 1000\, M_\odot$ where a small number of OB stars are present, almost exclusively as energetic binaries occupying the centre \citep{San12}. Dynamical interactions can then rid the cluster of OB stars on Myr timescales \citep{Pfl06}, ultimately leading to multiple discrete ages in the stellar population. 
 
 \subsection{Observational summary}
 \label{sec:obssummary}
 Some key features of the star and PPD population in the ONC that a successful model should reproduce are as follows:
\begin{itemize}
    \item The central density of the ONC is $n_0\approx 2 \times 10^4$~stars~pc$^{-3}$.
    \item Mass segregation suggests an epoch of cold collapse of the stellar population. 
    \item Kinematics data indicates that the stellar population is approximately in virial equilibrium without any bias between radial and azimuthal velocity components. 
    \item Youngest stars are preferentially found in the central regions (age gradient with radius). 
    \item Disc survival fractions are high ($\sim 80\%$) throughout the ONC. 
    \item The relationship between disc mass and stellar mass is found to be much more shallow than in other star forming regions (considerably sub-linear). 
    \item At the present day some proplyds exhibit mass loss rates due to FUV induced winds $\gtrsim 10^{-7}\, M_\odot$~yr$^{-1}$.
\end{itemize}
In the following sections we will address each of these points in turn to establish a model which explains both the stellar properties and the PPD population.
 
 \section{Numerical Method}
 
 \label{sec:nummethod}
 
Our modelling approach is similar to that employed in \citet{Win19}, where we presented an $N$-body model consistent with the observed morphology and kinematics for the stellar population in Cygnus OB2 and coupled it with disc evolution calculations. For the $N$-body integration we use \textsc{Nbody6++}, which has a built in prescription for a imposing a central potential on star particles. We emphasise that our models are simplified such that the physical parameters we describe have clear definitions and we are able to draw direct comparisons between simulations. They are not intended to capture the full complexity of the region, but rather to understand how the star formation history can influence the PPD population.
 
 \subsection{Stellar dynamical model}
 
 A key feature of our model is that we include three distinct stellar populations of different ages, as suggested by the findings of \citet{Bec17}. This is partly to make predictions on the PPD properties such that this conclusion can be tested, but also to simplify our models. We are primarily interested in the differential dynamical evolution of the older and younger stellar populations as gas is depleted, rather than whether or not star formation was continuous. We therefore impose a background gravitational potential on the star particles in the $N$-body simulations that represents the contribution from the reservoir of star forming material. To simplify the model, we will assume that the density of gas from which these stars form is spherically symmetric and the radial profile does not change over time except to be scaled by a factor due to mass loss. For numerical convenience, we assume the gas is distributed in a Plummer sphere, such that the gas density at time $t$ is:
 \begin{equation}
 \label{eq:plummerdense}
     \rho_\mathrm{g}(r,t) = \frac{3M_\mathrm{g}(t)}{4\pi a^3} \left( 1 + \frac{r^2}{a^2} \right)^{-5/2} =  \rho_\mathrm{g,0}(t) \left( 1 + \frac{r^2}{a^2} \right)^{-5/2}
 \end{equation} where $r$ is the radius within the cluster, $M_\mathrm{g}$ is the total gas mass with $\rho_{\mathrm{g},0}$ the central gas density, and $a=0.4$~pc is the scale radius (to give core radius $r_\mathrm{core} \approx 0.25$~pc -- physically this choice is of little importance since the stars in our model undergo an initial epoch of violent relaxation). We define $M_\mathrm{g}(t)$ to be a step-function, such that at discrete times a quantity of gas is transformed into stars. For three distinct stellar populations with masses $M_*^{(1)}$,  $M_*^{(2)}$ and  $M_*^{(3)}$, born at times $0$, $T_*^{(2)}$,  and $T_*^{(3)}$ respectively we have:
 \begin{equation}
 \label{eq:mgas}
M_\mathrm{g}(t) = \begin{cases}
 \left(M_*^{(2)} +M_*^{(3)}\right)/\epsilon_\mathrm{eff} &0<t<T_*^{(2)}  \\
 M_*^{(3)}/\epsilon_\mathrm{eff} &T_*^{(2)} \leq t<T_*^{(3)} \\
0&  T_*^{(3)} \leq t
\end{cases}
\end{equation} where $0<\epsilon_\mathrm{eff}\leq1$ is the effective star formation efficiency (SFE). The value of $\epsilon_\mathrm{eff}$ is the fraction of gas present in our initial conditions that will be converted into stars. We assume that $\epsilon_\mathrm{eff}$ is constant in both space and time for a given model, and adopt values of $\epsilon_\mathrm{eff}=0.1$, $0.3$ and $0.7$ to test the influence of SFE on our results. The lowest value $\epsilon_\mathrm{eff}=0.1$ yields a large initial gas mass that would be comparable to the mass of the entire Orion A filament \citep{Lom14}, which is physically implausible. We nonetheless consider this case to illustrate the influence of low SFE on the star and PPD populations. The gas mass evolution in equation~\ref{eq:mgas} is simplified; for example, the available star forming material has likely been replenished by inflowing gas \citep{Hac17}. Our model is appropriate if the rate of change of gas mass (including inflows, outflows and loss to stellar mass) is rapid during the discrete epochs of star formation and slow otherwise. While this may not be the case, in the absence of constraints on historic gas flows (or a fully hydrodynamical model), any alternative prescription would be arbitrary. For the ages of the populations, we assume the oldest formed $2.8$~Myr ago, with $T_*^{(2)} =1$~Myr and $T_*^{(3)}= 2$~Myr, consistent with the ages derived by \citet{Kro18}. This few Myr timescale for gas removal is also broadly consistent with the numerical findings of \citet{Ahm19} for a star of mass $34\,M_\odot$ in a molecular cloud with $10^3\, M_\odot \lesssim M_\mathrm{g}\lesssim 10^4 M_\odot$.

The spatial distribution of the stars, which are added to the simulation at intervals, follows the gas Plummer profile. We define the total stellar masses for each population: $M_*^{(1)}=  2290\, M_\odot$,   $M_*^{(2)}=  1140\, M_\odot$,   $M_*^{(3)}=  570\, M_\odot$, which is approximately consistent with the observed relative masses of the populations \citep{Bec17}. Each population $k$ is added at time $T_*^{(k)}$ (with $T_*^{(k)} = 0$), at which point the appropriate quantity of mass is removed from the gas potential according to equation~\ref{eq:plummerdense}.

We draw stellar masses $m_*$ from a \citet{Kro01} IMF:
\begin{equation}
\label{eq:imf}
\xi(m_*) \propto \begin{cases}
               m_*^{-1.3} \quad \mathrm{for } \, 0.08 \, M_\odot \leq m_*< 0.5 \, M_\odot\\ 
               m_*^{-2.3} \quad \mathrm{for } \, 0.5 \, M_\odot \leq m_* < 10 M_\odot \\
              0 \qquad \quad \, \mathrm{otherwise}
            \end{cases}
\end{equation} and assign them to positions at random. We do not consider primordially mass segregated initial conditions, except to place a star of mass $35 \, M_\odot$ (analogous to $\theta^1$C) at the closest position to the centre. Equation~\ref{eq:imf} is truncated above $10 \, M_\odot$ when drawing the remainder of the stellar masses. The reason for choosing such an IMF is that we aim to reproduce the unusual distribution of observed stellar masses that may have resulted from ejection of the most massive cluster members by dynamical interactions \citep{Pfl06,Kro18}. For modelling simplicity, and because we wish to prescribe the ages of our population to match observations, we do not directly simulate the ejection of massive stars by binary interaction and do not include binaries in our initial conditions. Instead, we consider a single massive star that occupies the cluster for its entire evolution. If the timescale of each of the discrete epochs of star formation is short ($\ll T_\mathrm{age}$, the overall age of the ONC), then this is approximately equivalent to a model in which massive stars are removed periodically and rapidly replaced during a new phase of star formation. Fixing the mass of the most massive star also has the benefit that we can draw direct comparisons of the PPD properties between models.

All stars are initially assigned a speed $v$ from a Maxwell-Boltzmann distribution:
\begin{equation}
    f(v) \propto v^2 \exp \left( {v^2} \right),
\end{equation} rescaled to give a 1D velocity dispersion $\sigma_v = 2$~km/s, close to the present day observed dispersion. We make this choice such that the stellar population is approximately in virial equilibrium with itself, however our models undergo cold collapse since they are subvirial with respect to the gas potential. The dynamical evolution is therefore only weakly sensitive to the initial velocity distribution and any initial substructure \citep{Par14}. The direction of the velocity vector is drawn from an isotropic distribution.

 \subsection{Interstellar flux}
 \label{sec:fluxcalc}
 To calculate the FUV and EUV flux experienced by the stars over their lifetimes we follow the method of \citet{Win19}. The total luminosity and effective temperatures of all stars with mass $>1\,M_\odot$ are taken from \citet{Sch92} for metallicity $Z = 0.02$ and the output closest to the time $1$~Myr. Atmosphere models by \citet{Cas04} give the wavelength-dependent luminosity, which we then integrate  over FUV ($6$~eV~$< h\nu < 13.6$~eV) and EUV ($h
 \nu> 13.6$~eV) photon energy ranges. The distances between stars is then tracked throughout the $N$-body simulations to give the flux experienced by each PPD. 
 
We additionally estimate the influence of interstellar extinction on the FUV flux. In order to do this, we integrate the gas density (equation~\ref{eq:plummerdense}) along the line of sight between two stars to give the surface density of gas between them:
\begin{equation}
    \Sigma^\mathrm{ext}_{ij}  = \left| \int_{\mathcal{C}\left(\bm{r}_i ,\bm{r}_j\right)} \rho_\mathrm{g}(|\bm{r}|) \, \mathrm{d} \bm{r} \right|
\end{equation} where $\bm{r}_i$ and $\bm{r}_j$ are the position vectors of stars $i$ and $j$ respectively and the vector $\bm{r}=\bm{0}$ at the centre of the Plummer potential. The path $\mathcal{C}$ is the straight line between them:
\begin{equation}
    \mathcal{C} \equiv \left\{ u \bm{r}_i + (1-u)\bm{r}_j : u \in [0,1] \right\}.
\end{equation} This is converted to an extinction estimating the ratio of FUV to visual extinction $A_\mathrm{FUV}/A_V = 2.7$ \citep{Car89}, and the column density of hydrogen required for $1$~mag extinction $N_\mathrm{H}/A_V = 1.8 \times 10^{21}$~cm$^{-2}$~mag$^{-1}$ \mbox{\citep{Pre95}}. By assuming a smooth Plummer gas density distribution we address the limit in which turbulent density fluctuations are on much smaller scales than the scale length of the cluster ($a=0.4$~pc). The opposite limit, for which the gas distribution is `clumpy' on large scales, corresponds to inefficient extinction of FUV photons. We will therefore bracket the limits in flux experienced by the PPD population by considering disc evolution with and without interstellar extinction. 

In the case of EUV photons, we must also consider the radius around the most massive star in which gas is ionised. This is the \citet{Str39} radius, which for a constant central local gas density $\rho_\mathrm{g,0}$ can be written:
\begin{equation}
\label{eq:RS}
R_\mathrm{S} \approx \left( \frac{3 \Phi_{\mathrm{i}} m_\mathrm{p}^2}{4\pi \alpha_\mathrm{B} \rho_\mathrm{g,0}^2} \right)^{1/3}  
\end{equation} where $m_\mathrm{p}$ is the proton mass, $\Phi_{\mathrm{i}}$ is the EUV photon count from the central star, and $\alpha_\mathrm{B} \approx 2.7 \times 10^{-13}$~cm$^3$~s$^{-1}$ is the recombination coefficient assuming a temperature $\sim 10^4$~K for the ionised gas. Even for the most modest non-zero central gas density in our model ($ \rho_\mathrm{g,0}\approx 2.1\times10^{-19}$~ g~cm$^{-3}$, for $1$~Myr~$<t<2$~Myr and $\epsilon_\mathrm{eff}=0.7$) this yields $R_\mathrm{S}\approx 0.03$~pc, where $\Phi_{\mathrm{i}} \sim 10^{49}$~photons~s$^{-1}$. Since $R_\mathrm{S}$ is small compared to the scale radius $a$, we consider (extincted) FUV photons only and disregard the influence of EUV photons on PPD evolution until the gas is completely removed (see Section \ref{sec:discmodel}). These assumptions do not contradict the present day $\sim$pc scale separation between the HII region and $\theta^1$C \citep[which lies in front along the line of sight --][]{Wen95}. In our model, the gas density within the ONC decreases as star formation events proceed, which would accelerate the D-type expansion of the ionisation front \citep[e.g.][]{Spi78}. Our assumption is that the EUV induced mass loss in PPDs only became efficient since the most recent star formation episode (although for the majority of discs, FUV photons still dictate mass loss rates in the externally driven wind).

 \subsection{Viscous disc model}
 \label{sec:discmodel}
 \subsubsection{Numerical method}
To calculate the surface density evolution of externally irradiated PPDs, we follow the method of \citet[][see also \citealt{Ande13, Ros17, Win19}]{Cla07}. A one dimensional viscous evolution model is calculated across a radial grid spread evenly $r^{1/2}$, for $r$ the radius within the disc. Viscosity is assumed to scale linearly with $r$, corresponding to a temperature which scales with $r^{-1/2}$ and a constant $\alpha$-viscosity parameter \citep[][see also \citealt{Har98}]{Sha73}. In isolation, this gives a surface density which evolves as:
\begin{equation}
\label{eq:sigmadisc}
\Sigma_\mathrm{disc}  = \frac{M_{\mathrm{disc},0}}{2\pi R_1^2 R} \exp\left(- \frac R T \right) T^{-1/5}
\end{equation} where $R_1$ is the scale radius, $M_{\mathrm{disc},0}$ is the initial disc mass, and $T$ and $R$ are scaled time $t$ and radius $r$ coordinates:
\begin{equation}
T\equiv 1+ \frac{t}{\tau_\mathrm{visc.}} \qquad R \equiv \frac {r}{R_1}
\end{equation} for a viscous timescale $\tau_\mathrm{visc.}$ at $R_1$ .

Mass is removed from the outer edge of the disc by the UV driven at a rate of 
$$
\dot{M}_\mathrm{wind} = \mathrm{max} \left\{ \dot{M}_\mathrm{FUV}, \dot{M}_\mathrm{EUV} \right\},
$$which ensures a sharp drop in surface density outside of some radius $R_\mathrm{disc}$. The FUV induced mass loss rates $\dot{M}_\mathrm{FUV}$ are interpolated from the \textsc{Fried} grid, which are computed over a range of stellar host mass, disc outer radius, disc mass and FUV flux \citep{Haw18b}. The EUV mass loss rate $\dot{M}_\mathrm{EUV}$  is calculated using the analytic estimate of \citet{Joh98}:
\begin{equation}
\label{eq:EUVloss}
\frac{\dot{M}_\mathrm{EUV} }{ M_\odot \, \mathrm{yr}^{-1}}= 5.8 \times 10^{-8} \left( \frac{d}{0.1\, \mathrm{pc}}\right)^{-1} \left(\frac{ f \Phi_\mathrm{i}}{10^{49} \, \mathrm{s}^{-1}} \right)^{1/2}  \left( \frac{R_\mathrm{disc}}{100 \, \mathrm{au}}\right)^{3/2},
\end{equation} where $d$ is the separation and $\Phi_\mathrm{i}$ is the ionising photon count from the radiating source, and $f$ is the fraction of ionising photons that penetrate the interstellar medium (we assume $f=1$ if gas has been expelled, $f=0$ otherwise -- see Section~\ref{sec:fluxcalc}).

 \subsubsection{Initial conditions}
 
Since we are partly interested in understanding the evolution of the host mass--disc mass (radius) relationship, we consider constant $M_{\mathrm{disc},0}= 0.1\, M_\odot$ ($R_{\mathrm{disc},0}=50$~au), independent of stellar mass. While such initial conditions may not represent the physical relationship between stellar mass and disc mass \citep[e.g.][]{Andr13,Pas16}, this choice allows us to clearly illustrate how the relationship steepens over time, since any correlation is due to depletion over the time-scale of the simulation. The influence of this assumption on our host mass dependent results are explored in Appendix~\ref{app:hostmass}, where we consider disc masses $M_{\mathrm{disc},0}= 0.1\, m_*$. 

We also aim to put an upper limit on the allowed value of $\alpha$, which along with $R_1$ sets the initial accretion rate 
\begin{equation}
\dot{M}_{\mathrm{acc},0} = \frac 3 2 \, \alpha M_{\mathrm{disc},0} \Omega_1 \left(\frac{ H_1}{R_1}\right)^2,
\end{equation} where $\Omega_1$ and $H_1$ are the angular velocity and scale height of the disc at $R_1$ respectively.  We assume $H_1/R_1 = 0.05$, and Keplerian rotation. We initially truncate the surface density at $R_{\mathrm{disc},0}$, such that the outer radius is well-defined (which is required to calculate the appropriate $\dot{M}_\mathrm{wind}$). We choose $R_1 = R_{\mathrm{disc},0}/2.5 = 20$~au, such that integrating the initial surface density (equation~\ref{eq:sigmadisc} with $T=1$) over radius gives $0.92\, M_{\mathrm{disc},0}$. We will investigate the dependence of our results on the assumed disc viscosity by varying $\alpha$.

 \section{Results and discussion}
 \label{sec:resanddisc}
 
The solution to the proplyd lifetime problem that we propose here is composite, such that it is not possible to point to a single physical consideration that reproduces the observed PPD sample in the ONC. In this section we sequentially review different observational constraints, linking them to our model and exploring the influence of parameter choices where necessary. Section~\ref{sec:stelldyn} is devoted to obtaining an appropriate dynamical model. In Section~\ref{sec:PPDevol} we explore the properties of the photoevaporating PPD population over time within the dynamical model we derive in Section~\ref{sec:stelldyn}. In this way we address each of the constraints summarised in Section~\ref{sec:obssummary}. For a reader interested only in the key considerations of our model, we summarise our proposed solution in Section~\ref{sec:PLP_sol}. 
 
 \subsection{Stellar dynamics}
 \label{sec:stelldyn}
 \subsubsection{Density distribution}
 
\begin{figure}
 \centering
      \includegraphics[width=0.45\textwidth]{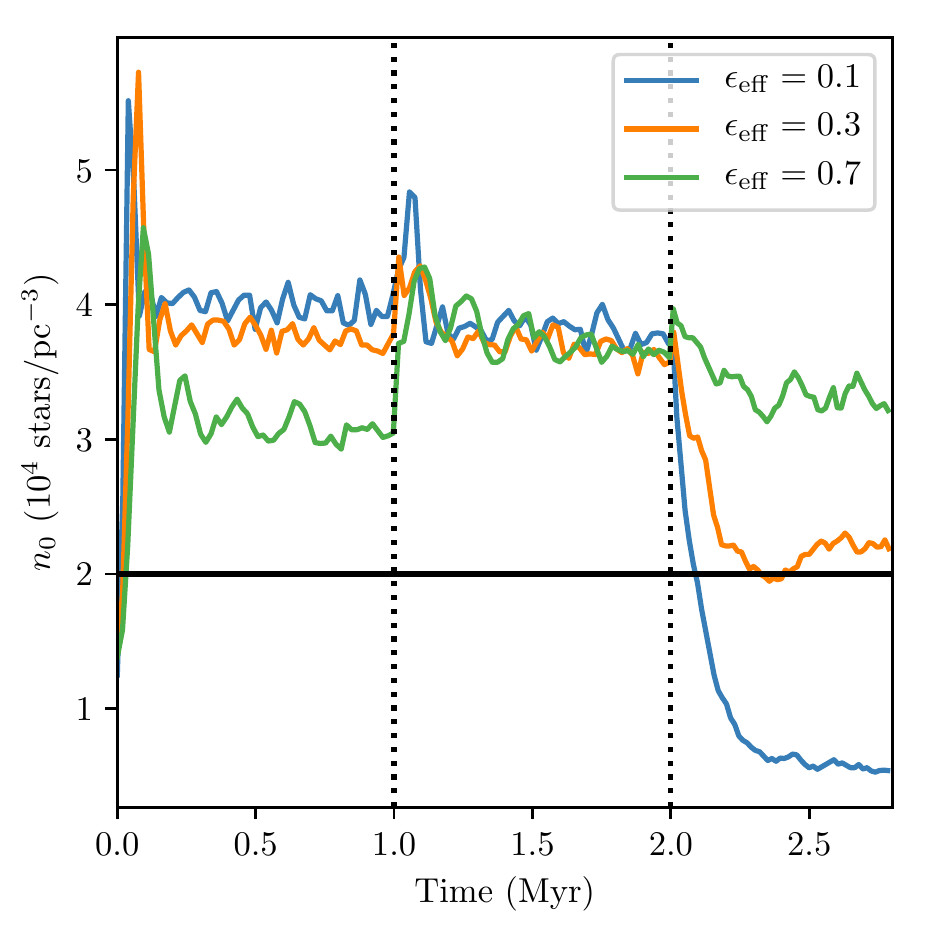}
     \caption{Average number density of stars $n_0$ inside of $0.2$~pc in each of the $N$-body simulations assuming varying effective SFE $\epsilon_\mathrm{eff}$. The black horizontal line is the central density as estimated by \citet{Hil97}. The vertical dotted lines are the times at which new stellar populations are introduced (and gas removed). }
     \label{fig:n0evol}
\end{figure}

 We initially consider how varying $\epsilon_\mathrm{eff}$ influences the stellar density and kinematics of our model. In general we expect low SFE to result in initially enhanced stellar density near the centre of the cluster due to the cold collapse of the stellar population (which, since we have imposed initial conditions that are virial with respect to the stellar component alone, is initially subvirial with respect to the stellar and gas mass). After gas expulsion, models with a lower $\epsilon_\mathrm{eff}$ expand more rapidly as the dynamical interactions accelerate stars in the core. This principle is demonstrated in Figure~\ref{fig:n0evol}, where the central stellar number density $n_0$ is shown as a function of time. We find that low $\epsilon_\mathrm{eff}$ initially increases $n_0$, which subsequently decreases after $2$~Myr, when all the gas is removed. Since the central density does not change rapidly over time by the end of the simulation, uncertainties in the age of the ONC do not translate into uncertainties in $\epsilon_\mathrm{eff}$. Thus, the present day $n_0$ represents a good test of a successful model, and a SFE of $\epsilon_\mathrm{eff} = 0.3$ reproduces the observed $n_0 \approx 2\times 10^{4}$~stars~pc$^{-3}$ \citep{Hil97}. This is broadly consistent with the observed SFE for embedded clusters on a $\sim 1$~pc scale \citep{Lad03}, and matches estimates for the core of the ONC \citep{Meg16}. We consider how the assumed SFE influences other observational constraints below.

\subsubsection{Kinematics} 
 
 \begin{figure}
 \centering
      \includegraphics[width=0.45\textwidth]{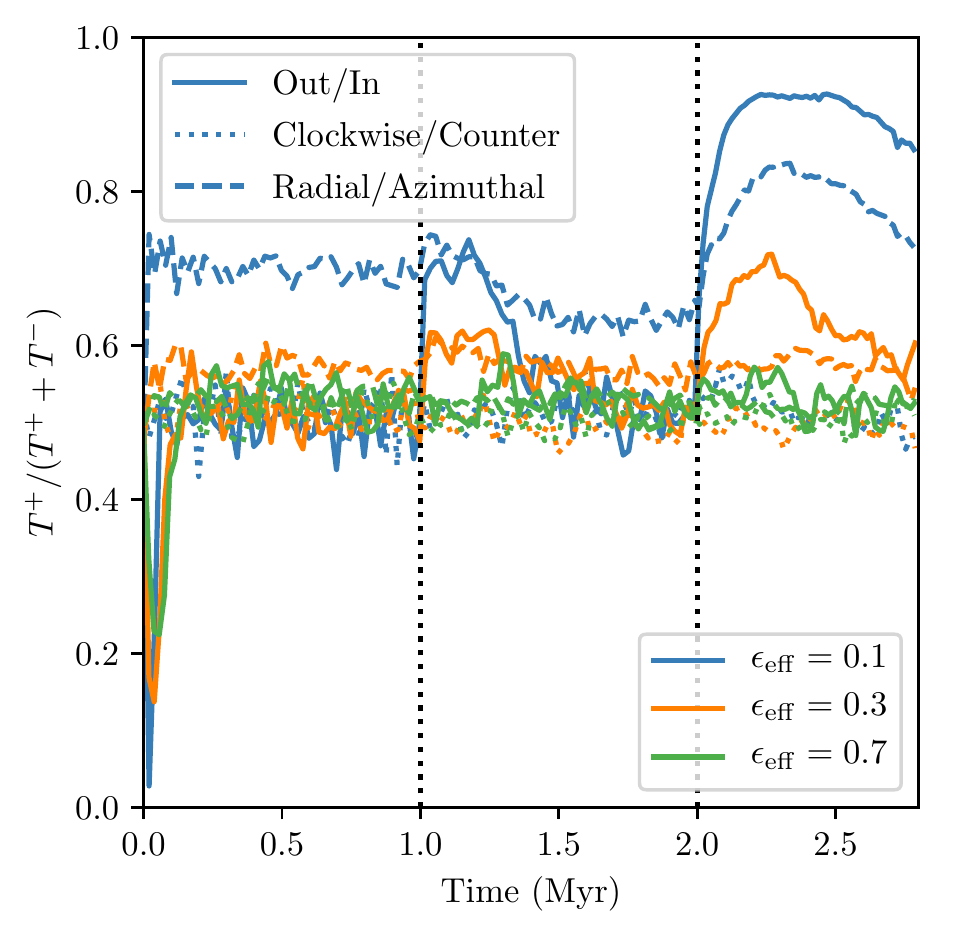}
     \caption{Ratio of the kinetic energy in each projected direction (radial and azimuthal) of the stellar population as a function of time. We show the split between outwards/inwards (solid), clockwise/counterclockwise (dotted) and radial/azimuthal (dashed) kinetic energy for each star formation efficiency $\epsilon_\mathrm{eff}$. Each model goes through an early phase of rapid collapse since the stellar components are initially subvirial with respect to the gas potential. For $\epsilon_\mathrm{eff}\lesssim 0.3$, a period of rapid expansion occurs after a star formation event (at times marked by vertical dotted lines). For $\epsilon_\mathrm{eff}=0.1$ rapid expansion continues until the present time.  }
    \label{fig:KE}
\end{figure}

No strong biases between radial and azimuthal velocities are apparent in the stellar kinemtics of the ONC, nor evidence of expansion in the radial components \citep{Kim19}. We show the split between kinetic energy $T$ in each direction (denoted $T^{+/-}$) in Figure~\ref{fig:KE}. No model exhibits bias between clockwise and counter  clockwise directions, as expected since we did not include any rotation in our initial conditions. All models also undergo a short, rapid phase of collapse early on, which is again expected since the initial conditions are subvirial with respect to the gas potential. For $\epsilon_\mathrm{eff}=0.1$ this collapse precedes a sustained phase of evolution where the stellar components are on radial orbits (the radial kinetic energies exceed azimuthal components). At each period of star formation (and gas expulsion) there is also a phase of rapid expansion, and the most recent episode continues until the end of the simulation. For $\epsilon_\mathrm{eff}=0.3$, their is little bias towards radial kinetic energy, and the most recent phase of expansion ends before $2.8$~Myr. This would be consistent with the observational constraints. 

 \subsubsection{Age segregation}

\begin{figure}
 \centering
       \includegraphics[width=0.45\textwidth]{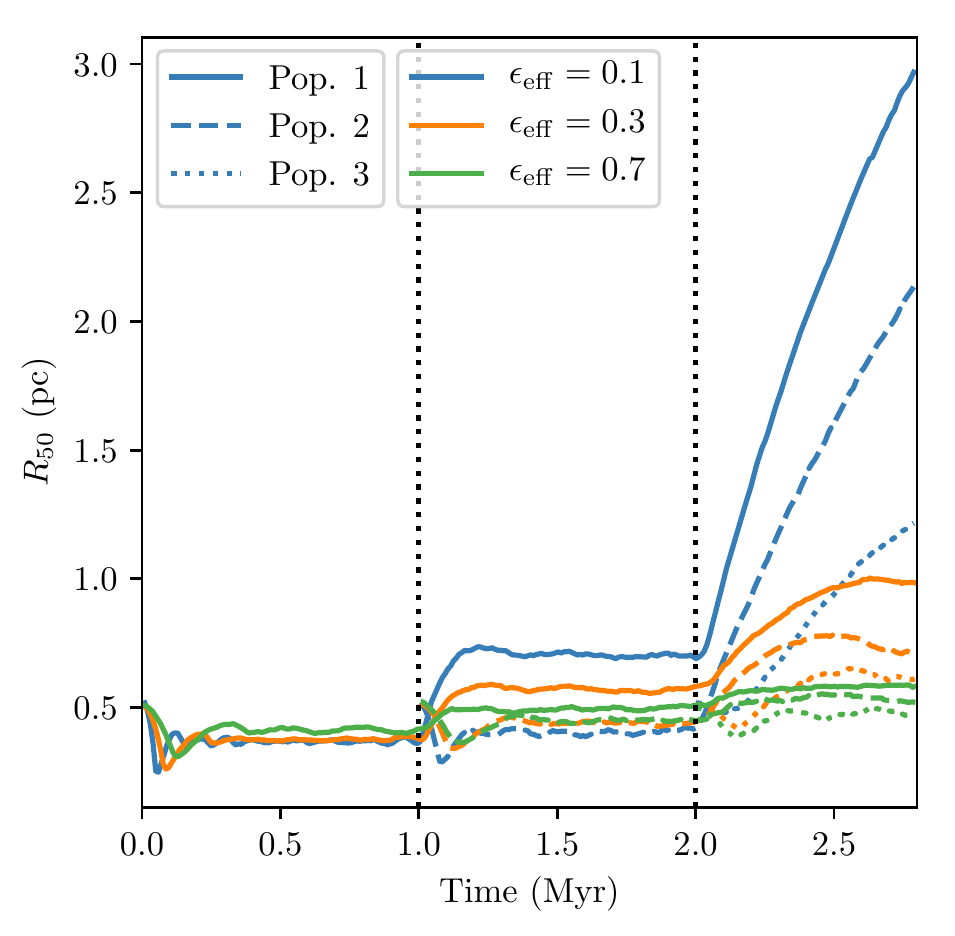}
     \caption{Half-mass radius ($R_{50}$) evolution for the three stellar populations in our $N$-body simulations with varying effective star formation efficiency $\epsilon_\mathrm{eff}$. The youngest stars (Pop. 3) are more centrally concentrated than the older populations by the present day (after $2.8$~Myr of evolution) for smaller values of $\epsilon_\mathrm{eff}$. The dotted vertical lines are the times at which each stellar population is born. }
     \label{fig:rhmpops}
\end{figure}

We aim to reproduce the finding that young stars are preferentially found towards the centre of the ONC. In Figure~\ref{fig:rhmpops} we show the half-mass radii of the different stellar populations in our models over time. The true half-mass radius of the ONC is dependent on its assumed definition (i.e. truncation radius) and membership. In addition our (Plummer) density profile is chosen for numerical reasons, and does not necessarily reflect the physical profile. We therefore do not aim to reproduce the absolute half-mass radius, but achieve a differential radial extent between the populations. We find that if SFE is too high, the older stars are not significantly accelerated by dynamical encounters in the core and therefore there is little segregation between populations by the present time (as is the case for $\epsilon_\mathrm{eff}=0.7$). For sufficiently low $\epsilon_\mathrm{eff} \lesssim 0.3$, the older stars (i.e. Pop. 1) are more extended than the younger stars (i.e. Pop. 3). However, if the SFE is too low, the older stars form an extended `halo' that rapidly expands outwards, as can be seen for the $\epsilon_\mathrm{eff}=0.1$ case. Since this is not supported by observations of the ONC, this suggests $\epsilon_\mathrm{eff}\approx 0.3$ is appropriate, as in the previous sections.

\subsection{PPD evolution}
\label{sec:PPDevol}
 \subsubsection{`Survival' fractions}
 \label{sec:PPDsurv}
\begin{figure}
 \centering
       \includegraphics[width=0.45\textwidth]{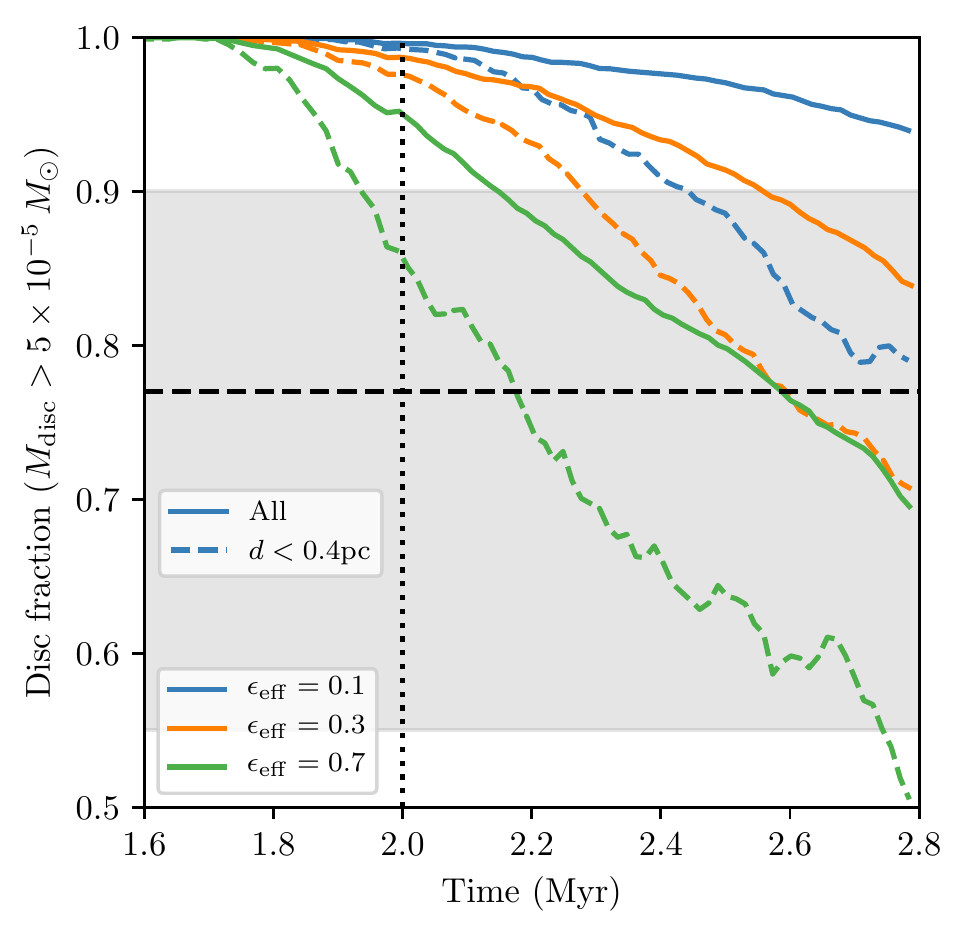}
     \caption{Surviving disc fraction in our simulations including efficient extinction as a function of time at various effective SFEs with $\alpha=10^{-3}$. Dashed lines are for stars with a projected distance from the centre $d<0.4$~pc, while solid lines are for all the stars within $d<3$~pc of the centre. The horizontal dashed line is the fraction of stars that host discs in the central $0.5$~pc$^2$ found by \citet{Lad00}, while the grey shading is the range of values compatible with the findings of \citet{Hil98b} across the entire ONC. The dotted vertical line is the time at which the last of the gas is converted into the Pop. 3 stars. }
     \label{fig:dfracs_sfe}
\end{figure}

 The majority of stars \citep[$55$--$90$\% across the entire region, and $\sim 77\%$ within $\sim 0.2$~pc of the centre --][]{Hil98b, Lad00} in the ONC exhibit an NIR excess associated with the presence of circumstellar material, with no clear evidence of spatial dependence \citep[see also][]{Jef11}. In Figure~\ref{fig:dfracs_sfe} we consider the dependence of the evolution of the disc survival fraction (i.e. discs with masses $M_\mathrm{disc}>5\times10^{-5}\, M_\odot$) on the SFE. We find that the surviving disc fraction decreases with increasing SFE, as expected since the influence of interstellar extinction is reduced at lower gas densities. The results for $\epsilon_\mathrm{eff}=0.3$, which best matched the stellar kinematics also reproduce the disc fractions in the central regions ($\sim 70\%$).
 
 This finding demonstrates one component of our proposed solution to the proplyd lifetime problem. This is the principle that a combination of interstellar extinction and an extended period of star formation, where stars are initially subvirial with respect to the gas, can result in enhanced disc fractions at the present day consistent with those observed in the ONC.
 
\begin{figure}
 \centering
       \includegraphics[width=0.45\textwidth]{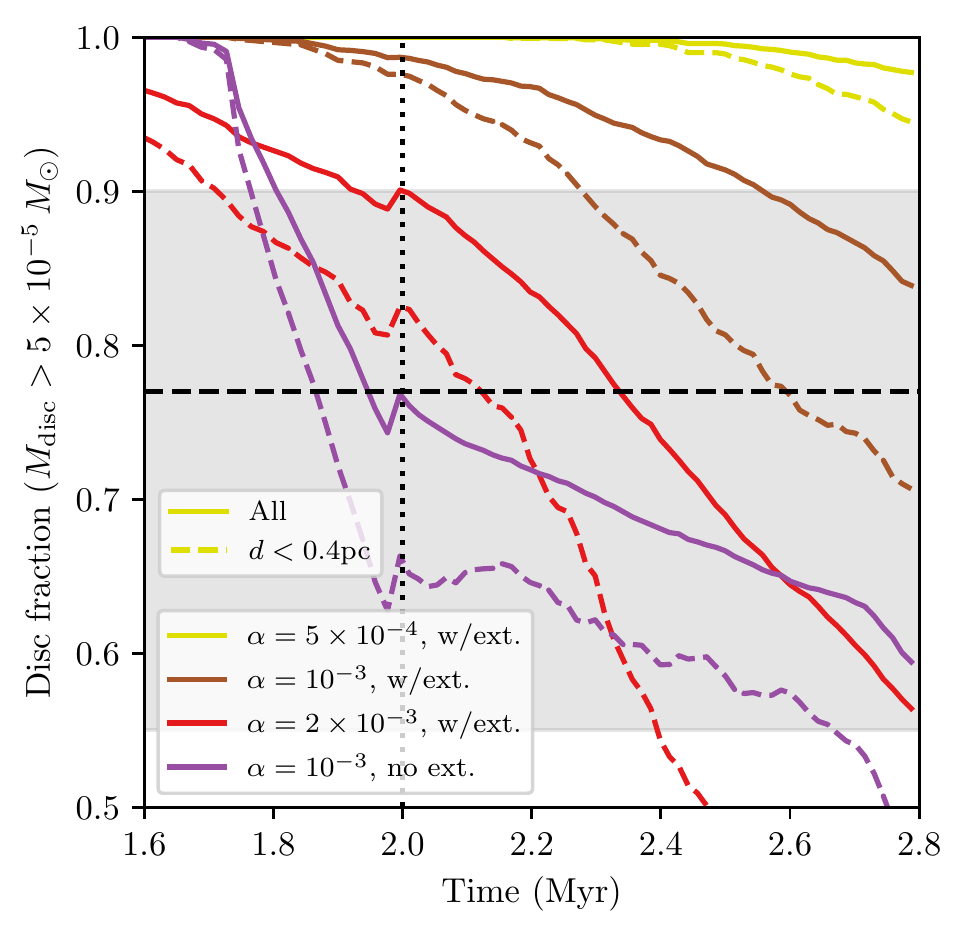}
     \caption{As in Figure \ref{fig:dfracs_sfe} except for fixed SFE $\epsilon_\mathrm{eff}=0.3$ and varying $\alpha$. We additionally show how neglecting extinction changes the disc fractions (purple lines).  }
     \label{fig:dfracs_ae}
\end{figure}

 We also consider the fraction of surviving discs in the core of the ONC (which experience the greatest UV flux) with respect to the broader population. The survival fractions within a projected distance $d<0.4$~pc of the centre are consistently $\sim 10$--$20\%$ lower than over the entire ONC across all values of $\epsilon_\mathrm{eff}$ in our simulation. A number of factors that we do not account for here may influence the differential between the core and outer disc fractions. Perhaps the most probable is the uncertainty or intrinsic variation in the ages of the respective stellar populations. It is also possible that the initial disc properties are variable over time; for example, younger discs may be initially less extended such that photoevaporative mass loss rates are initially lower. Additionally, the core gas densities we have assumed throughout the embedded phase are $\sim 10^{-19}$--$10^{-17}$~g~cm$^{-3}$, such that ram pressure can truncate a disc down to $<50$~au within $1$~Myr \citep[although this conclusion depends on geometry and the relative velocity of the ISM -- see][]{Wij17a}. Stellar densities in the core of $\sim 4\times 10^4$~pc$^{-3}$ persist for the first $2$~Myr in our model (Figure~\ref{fig:n0evol}), such that encounter driven disc truncation is possible. However, mean tidal truncation radii are still $>50$~au on this timescale \citep{Win18b} and ram pressure stripping (or accretion) is dominant over dynamical encounters \citep{Wij17b}. We do not explore these many possibilities here because of the large parameter space and uncertainties make computing the disc properties for each possible scenario impracticable and unhelpful. The intention of our model is rather to demonstrate a feasible mechanism by which discs near the core of the ONC can survive to the present day. 
 
However, the assumptions in our disc evolution model (in particular $\alpha$ and the extinction properties), also alter the surviving disc fraction.  We test the influence of varying these parameters in Figure~\ref{fig:dfracs_ae}. In general, higher $\alpha$ and the absence of extinction speed up disc dispersal, as expected. For $\alpha<10^{-3}$ the disc fraction begins to saturate close to $100\%$, leading to a much smaller differential disc fraction between the inner and outer regions. Decreasing the efficiency of extinction has a similar influence on the differential disc fraction, in this case hastening disc dispersal in older (more spatially extended) populations while having no influence on the youngest stars that are born after the gas has been removed. Since the disc fraction throughout the ONC is observed to be high, the low viscosity models including extinction are favoured. Hereafter, we will label the model including extinction and with $\alpha= 10^{-3}$ (for $\epsilon_\mathrm{eff}=0.3$) our `fiducial model'.  

Although disregarding interstellar extinction does decrease the survival fraction across the entire ONC, we highlight that a significant number of PPDs still survive in this case \citep[c.f.][where the authors find that no discs survive for reasonable initial masses after a few Myr]{Sca01}. This illustrates another key element of the solution to the proplyd lifetime problem: the evolution of the disc outer radius. Previous studies have assumed that the disc radii remain constant over time, whereas here we have factored in the depletion of the disc from the outer edge. External photoevaporation becomes less efficient at smaller radii, where material is more strongly gravitationally bound to the host star. Depending on the accretion rate, a disc can therefore survive in regions that are subject to strong UV fields for a longer period than predicted by models that do not take into account outer radius depletion.  

A number of other considerations could influence the overall disc fraction in the ONC. For discs in the outer regions (experiencing more modest FUV irradiation by $\theta^1$C than those in the core), depletion may be hastened by internal photoevaporation \citep[e.g.][]{Ale06, Ale06b, Wan17}. Binary companions may also influence dispersal rates \citep[e.g.][]{Ale12,Har12,Kra12,Ros18}. Factoring in such considerations is beyond the scope of this work, but could result in a modest change in the appropriate value for $\alpha$ required to match the disc fraction in the ONC.

  \subsubsection{Mass distribution}
  
\begin{figure}
 \centering
       \includegraphics[width=0.45\textwidth]{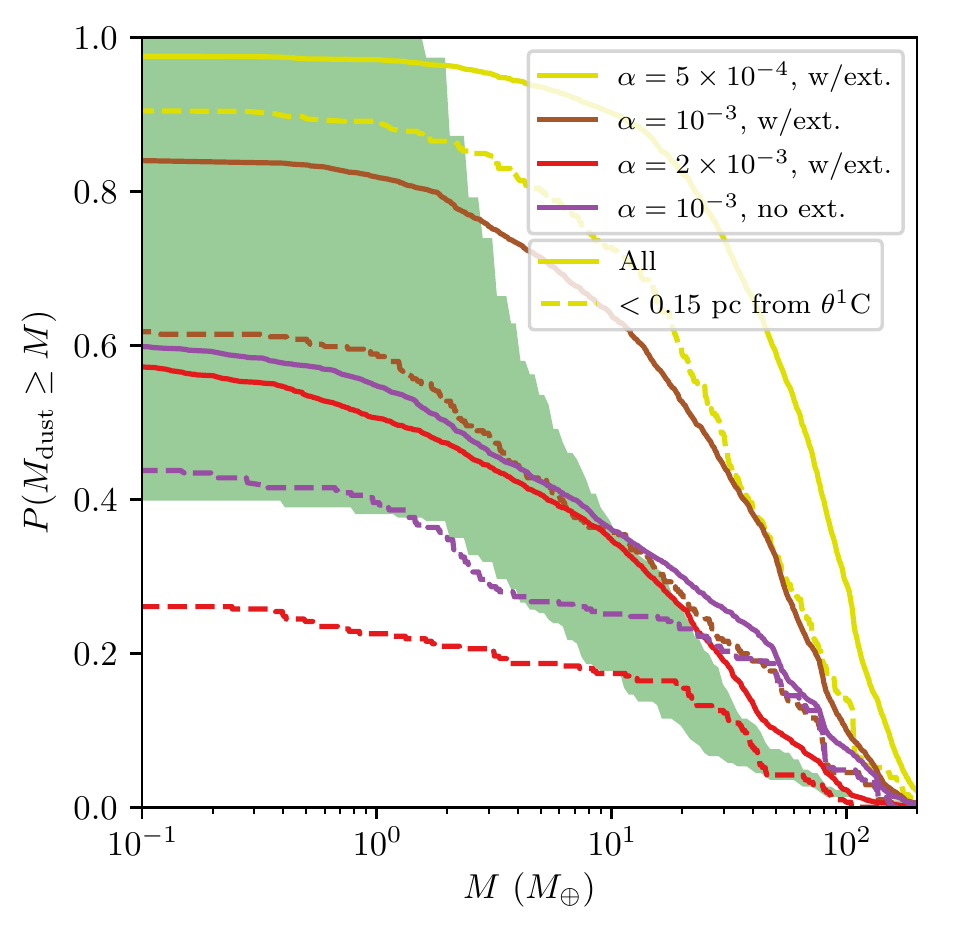}
     \caption{Cumulative distribution of PPD dust masses (assuming dust-to-gas ratio $10^{-2}$) after $2.8$~Myr of evolution for discs within $0.15$~pc in projected distance from $\theta^1$C. We show results for the fiducial model ($\epsilon_\mathrm{eff}=0.3$, $\alpha =  10^{-3}$ and including efficient extinction), and comparisons with models with higher $\alpha = 2\times 10^{-3}$, lower $\alpha=5\times 10^{-4}$ and where extinction effects are ignored. The green shaded region corresponds to the range compatible with the constraints in \citetalias{Eis18} for the core of the ONC (to be compared with the dashed lines), assuming the dust properties discussed in the text.}
     \label{fig:mdcf}
\end{figure}

  When we consider disc properties we will draw comparisons to the findings of the recent \citetalias{Eis18} investigation of dust masses in the core of the ONC, which are depleted compared to those in the surrounding OMC2 \citep{vTer19}. There are many sources of systematic uncertainty inherent in such observations, that are discussed in that study. These include, but are not limited to, assumed dust-to-gas ratio $\Sigma_\mathrm{dust}/\Sigma_\mathrm{gas}$, opacity $\kappa_\nu$ and temperature $T_\mathrm{dust}$. One other important consideration that we highlight here is the heterogeneous sensitivity across the central region due to the variable background emission, particularly from the BN/KL and OMC1 regions. This leads to variable upper flux limits on non-detections within the sample. This effect was not taken into account in the analysis of \citetalias{Eis18} and therefore we repeat the construction of the cumulative dust mass distribution, which we show in Figure~\ref{fig:mdcf}. We collect the measured fluxes and $4$-$\sigma$ upper limits (non-detections) of \citetalias{Eis18}, and compute upper and lower limits of the distribution by assuming maximal and minimal fluxes for the non-detections. For consistency, we follow \citetalias{Eis18} in calculating the dust mass using the equation:
  \begin{equation}
  M_\mathrm{dust}  = \frac{S_{\nu, \mathrm{dust}} D^2 }{\kappa_\nu  B_\nu(T_\mathrm{dust})},
  \end{equation} where $D=414$~pc is the distance to the ONC, $B_\nu$ is the Planck function and $T_\mathrm{dust}=20$~K. The opacity is taken to be:
  \begin{equation}
  \kappa_\nu =\kappa_0 \left( \frac{\nu}{\nu_0}\right)^{-\beta},
  \end{equation} where $\kappa_0 = 2.0$~cm$^2$~g$^{-1}$ at a wavelength of $1.3$~mm, and $\beta=1.0$ \citep{Hil83, Bec90}. In Figure~\ref{fig:mdcf} we show the upper and lower limits in the dust mass distribution accounting both for the quoted uncertainties and the spatially varying sensitivity (shaded region).

 The mass distribution we obtain from our models is dependent on the choice of $\alpha$ and the degree of extinction. In the majority of our simulations we find that more massive discs (with dust masses $\gtrsim 50\, M_\oplus$) than are inferred from the observations are present at the present day. We highlight that, since many of the discs in the ONC are compact (see Section~\ref{sec:discradii}), higher mass PPDs could become optically thick such that the mass estimates are lower limits. The difference could also be due to our adopted disc mass distribution. This would not be surprising since we fix all discs at a mass of $M_{\mathrm{disc},0}=0.1\, M_\odot$. Physically we might expect some distribution of initial disc masses and the mean value may be lower, especially considering the majority of stars in the IMF have a mass $m_*<1\, M_\odot$ such that $M_{\mathrm{disc},0}$ approaches the gravitational stability limit. However, this choice does allow us to effectively rule out models with $\alpha \gg 10^{-3}$, since $\alpha =2\times10^{-3}$ results in a lower disc fraction (Figure~\ref{fig:dfracs_ae}) and fewer discs with mass $\gtrsim 1\, M_\oplus$ (Figure~\ref{fig:mdcf}) than is consistent with observations.

  \subsubsection{Mass vs. stellar mass}
  \label{sec:MdiscvMstar}
 Since the sensitivity limit for the dust masses in the \citetalias{Eis18} sample is largely set by the background emission (except in $<10\%$ of cases where free-free, which may originate in the disc or background large scale flows, dominates at $850\, \mu$m), there is no reason why the stellar mass of the host should correlate with the sensitivity (although, since stellar masses are not quoted for the non-detections in \citetalias{Eis18}, it is not easy to assess the effect of possible correlations between stellar mass and level of background emission). The flat host mass--disc mass relationship found in that study ($M_\mathrm{disc} \propto m_*^\beta$ where $\beta=0.25 \pm 0.15$) is therefore likely to be physical in origin. In this section we compare this with the PPD mass dependence on host mass we obtain from our model.
 
\begin{figure*}
     \subfloat[\label{subfig:msmd_5e-4} $\alpha=5\times 10^{-4}$ ]{\includegraphics[width=0.45\textwidth]{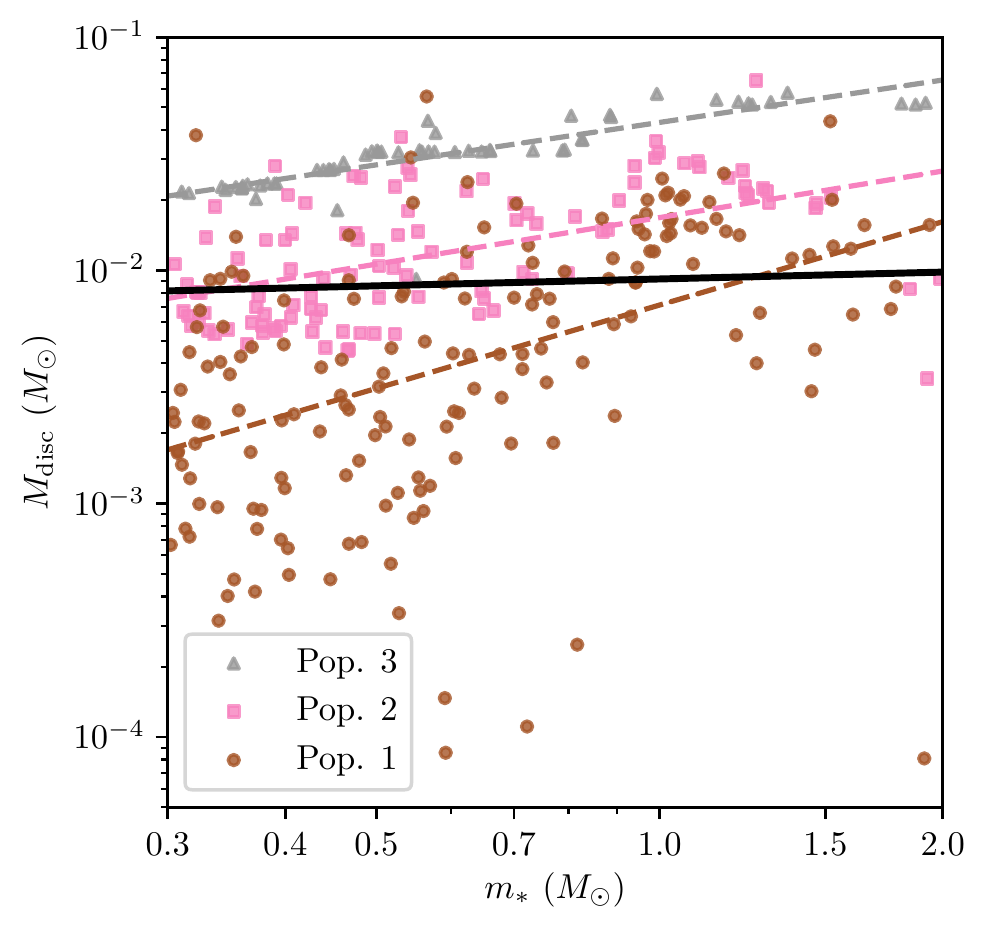} 
     }
       \subfloat[\label{subfig:msmd_1e-3} $\alpha=10^{-3}$ (`fiducial model')]{  \includegraphics[width=0.45\textwidth]{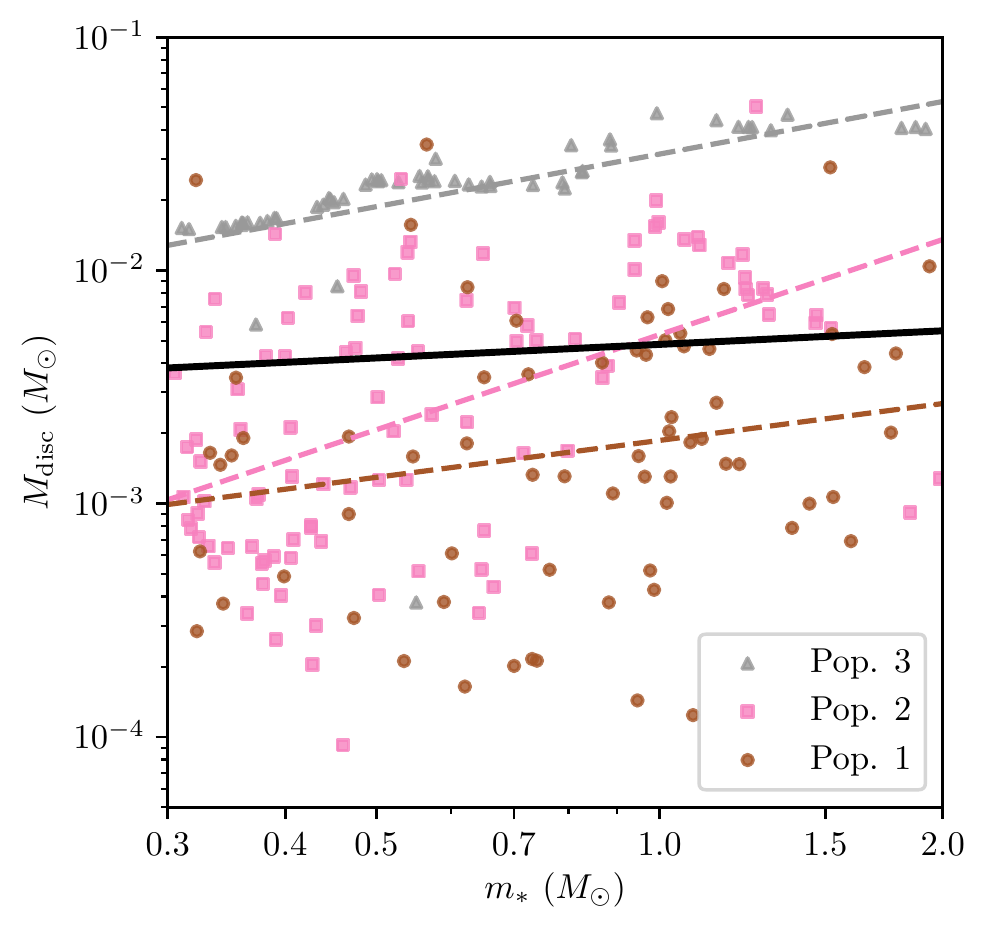}}
     \caption{Disc mass $M_\mathrm{disc}$ as a function of stellar host mass $m_*$ after $2.8$~Myr of evolution for an $\epsilon_\mathrm{eff}=0.3$ simulation including interstellar extinction. We show results for two different viscous parameters, $\alpha=5\times 10^{-4}$ (Figure~\ref{subfig:msmd_5e-4}) and $\alpha = 10^{-3}$ (Figure~\ref{subfig:msmd_1e-3}). The fitted power laws ($M_\mathrm{disc} \propto m_*^\beta$ have slopes with $\beta = 0.10$, $0.20$ respectively. \citetalias{Eis18} found $\beta = 0.25 \pm 0.15$. We have also shown the relationships that would be obtained in our model taking only a single population of stars as dashed lines. For $\alpha=10^{-3}$ ($5\times 10^{-4}$) these are $\beta =0.52$ ($1.19$), $1.35$ ($0.66$) and $0.75$ ($0.60$) for Pop. 1, Pop. 2 and Pop. 3 respectively. The shallower dependence for Pop. 1 stars in the $\alpha=10^{-3}$ case originates from a decreased number of discs surviving around lower mass stars (preferentially low mass discs).} 
     \label{fig:mhostmdisc}
   \end{figure*}

In Figure~\ref{fig:mhostmdisc} we show the distribution of the total disc mass $M_\mathrm{disc}$ as a function of stellar mass $m_*$. Similarly to \citetalias{Eis18} we find that, when all discs are considered, the disc mass is only weakly correlated to the host star ($\beta=0.10$, $0.20$ for $\alpha=5\times 10^{-4}$, $10^{-3}$ respectively). However, this is due to the fact that each population introduces contamination into the sample such that the correlations is `washed out' by the stellar age differential. In fact, when we decompose the populations and fit power laws individually, we do find a stronger correlation (even for Pop.~3 stars). In general this relationship steepens over time, with Pop. 3 stars exhibiting a much stronger correlation than Pop.~2 stars in both the $\alpha=5\times 10^{-4}$ and $\alpha=10^{-3}$ cases. However, in the latter case (Figure~\ref{subfig:msmd_1e-3}) the Pop.~1 stars again demonstrate a weaker correlation. This is because a higher fraction of discs around low mass stars have been dispersed in this case, and are therefore not included in the power law fit. This problem would be compounded by a poor sensitivity limit in observations.

   Our findings indicate a clear empirical test in terms the stellar population definitions in \citet{Bec17} by correlating stellar age with PPD properties. Additionally, the shallow relationship found by \citetalias{Eis18} is at odds with that found in other regions \citep[e.g.][]{Andr13,Pas16}. The mechanism we present here represents a way to reconcile these findings in the context of other PPD populations. 
   
   We must also investigate the possibility that our findings of a weak correlation between $M_\mathrm{disc}$ and $m_*$ is a result of our initial conditions since we have assumed they are initially independent. This assumption is explored in Appendix~\ref{app:hostmass} where we assume $M_{\mathrm{disc},0}= 0.1 m_*$. In this case we find a similar weak correlation across all populations ($\beta=0.28$), but steeper power law relationships within a single population. This confirms that the multiple stellar populations or extended period of star formation can explain the apparent lack of correlation between $M_\mathrm{disc}$ and $m_*$ for a range of initial conditions.

 \subsubsection{Radii vs. stellar mass}
 \label{sec:discradii}
\begin{figure}
 \centering
       \includegraphics[width=0.45\textwidth]{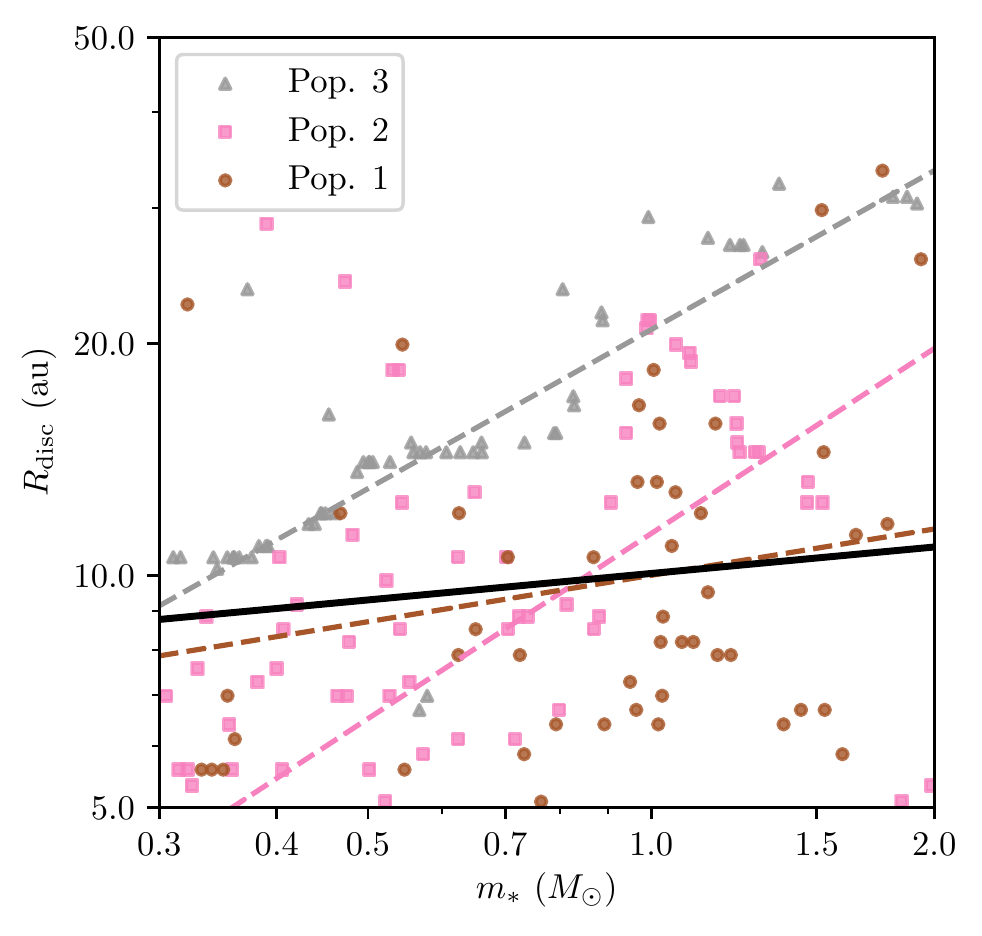}
     \caption{Disc outer radius $R_\mathrm{disc}$ as a function of stellar host mass $m_*$ in our fiducial model after $2.8$~Myr. The red line fit to the results follows $R_\mathrm{disc} = R_0 \, (m_* / 1\, M_\odot)^\gamma$ where $R_0 = 10.4$~au and $\gamma = 0.114$. These are consistent with the values $R_0\sim 10$--$20$~au and $\gamma = 0.15\pm 0.07$ found by \citetalias{Eis18}. For each population (dashed lines) $\gamma = 0.20$, $0.79$ and $0.68$ for Pop. 1, 2 and 3 respectively. }
     \label{fig:discrads}
\end{figure}

 Disc radii, while challenging to measure, are in some respects a superior probe of the influence of photoevaporation than stellar mass. For mass loss rates $\gtrsim 10^{-7} \,M_\odot$~yr$^{-1}$, PPDs can be rapidly depleted down to $\sim R_g$ (or the radius at which viscous spreading balances mass loss in the wind). However, this does not circumvent the problem posed by multiple stellar populations, which complicates attempts to fit a power law relationship with stellar mass. This is shown in Figure~\ref{fig:discrads}, where the relationship we fit to the whole population is flat, consistent with that found by \citetalias{Eis18}. We also show disc radii as a function of stellar mass, decomposed into the individual stellar populations. In this case, $R_\mathrm{disc} (m_*)$ does not significantly steepen with time, and we would therefore expect similar power law relations between stellar populations. We do however find an offset on the relationships, with older stars having less extended PPDs. As before, we also find that for the Pop. 1 stars, an attempt to fit a power law to the surviving discs again produces a flatter relationship. This is because discs around low mass stars only survive if they experience lower FUV field strengths, and thus have larger extents.
 
\subsubsection{Mass loss rates and ionisation fronts}

\begin{figure*}
 \centering
      \subfloat[\label{subfig:msmdot} \citetalias{Eis18} field of view.]{ \includegraphics[width=0.45\textwidth]{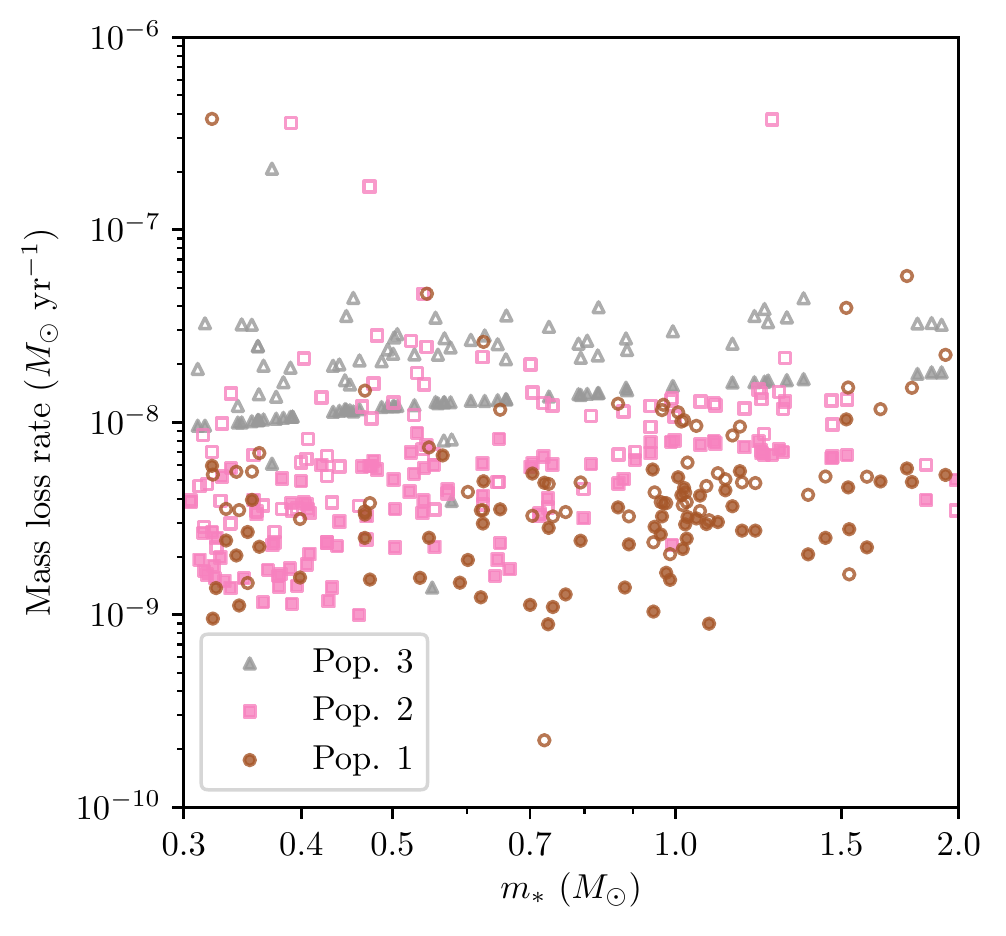}}
      \subfloat[\label{subfig:dmmdot} Variable distance to the $\theta^1$C analogue.]{ \includegraphics[width=0.45\textwidth]{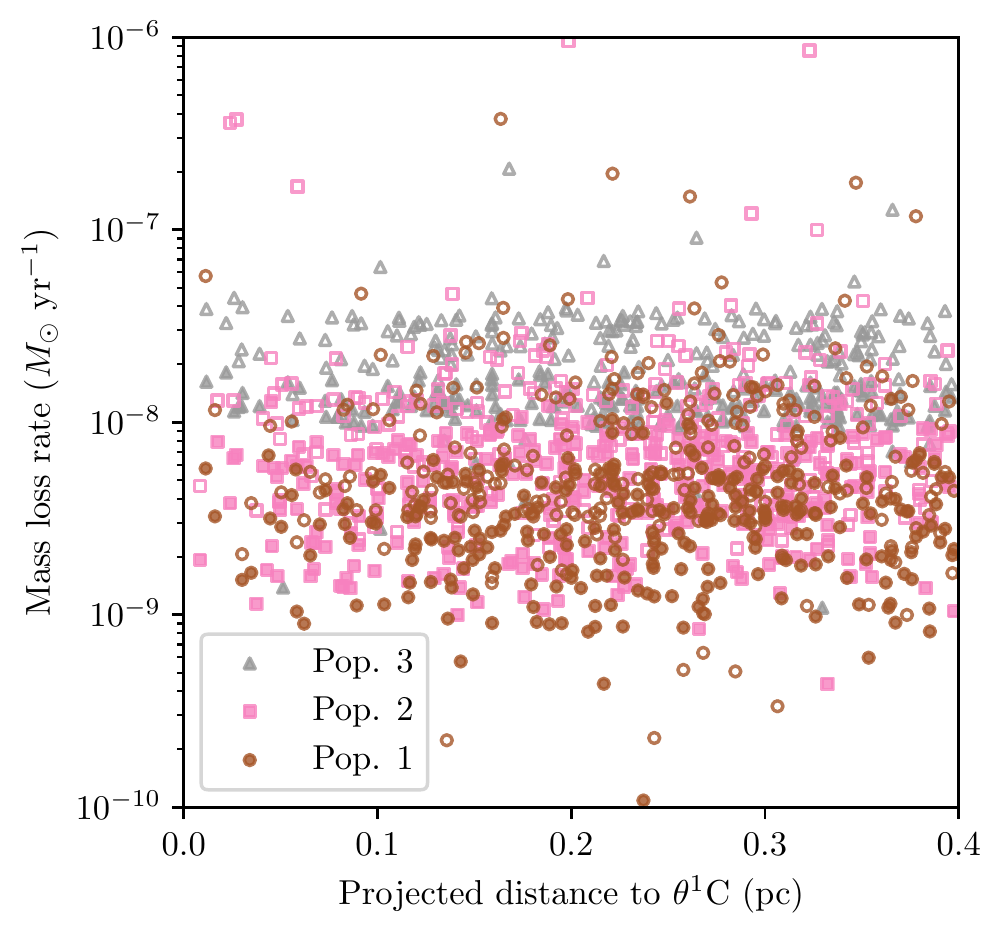}}
     \caption{Mass loss rates of the PPD populations. The loss rate in the photoevaporative winds (calculated using the \textsc{Fried} grid -- \citealt{Haw18b}) are shown in hollow shapes, while the accretion rates onto the host stars are shown as solid colours. In Figure~\ref{subfig:msmdot} we show the mass loss rates for the sample shown in the previous disc samples in the \citet{Eis18} field of view. In Figure~\ref{subfig:dmmdot} we show mass loss rates as a function of projected distance from the most massive star in our simulation.}
     \label{fig:mdot}
\end{figure*}

 Another important physical constraint is the mass loss rates, both due to the UV photons $\dot{M}_\mathrm{wind}$ and accretion $\dot{M}_\mathrm{acc}$. We show these values as a function of stellar mass at the end of our simulation in Figure~\ref{subfig:msmdot} for the sample in the \citetalias{Eis18} field of view. The most immediately obvious finding is that the distributions of both $\dot{M}_\mathrm{wind}$ and $\dot{M}_\mathrm{acc}$ are nearly independent of stellar mass. We interpret this to be due to the initial rapid dispersal of discs around low mass stars down to smaller radii than high mass stars (as discussed in Section~\ref{sec:discradii}). All discs then reach a similar total mass loss rate $\dot{M}_\mathrm{disc}$, but for lower mass stars this is equivalent to a shorter instantaneous dispersal timescale ${M}_\mathrm{disc}/\dot{M}_\mathrm{disc}$. Fluctuations in the UV flux experienced by these stars mean that mass loss rates can sometimes be much larger. In one scenario, a disc that has never previously experienced strong UV fields may migrate inwards and suddenly experience rapid mass loss. Alternatively, a disc that has been previously exposed to strong fields viscously expands in regions of low flux as it migrates outwards, only to be rapidly depleted again when it migrates back to the central regions. The latter mechanism is more efficient for low mass stars for which the gravitational radius $R_{g}$ is smaller, and the viscous timescale at $R_{g}$ is therefore shorter. This means that the disc can re-expand rapidly when it migrates outwards, exposing itself to higher mass loss rates upon its return to the centre. 
 
 For low mass stars ($m_* \lesssim 0.7 \, M_\odot$) we also find a dearth of discs for with significant $\dot{M}_\mathrm{wind}>10^{-10}\, M_\odot$~yr$^{-1}$ in the Pop. 1 sample. This can also be interpreted in terms of the radius evolution of the PPDs. As discussed in Section~\ref{sec:discradii} (and evident from Figure~\ref{fig:discrads}), the only Pop. 1 PPDs which survive with extents $\gtrsim R_g$ around low mass stars and apparent separations $\lesssim 0.15$~pc from $\theta^1$C are those which lie at greater physical (3D) separations. The exceptions to this principle are individual stars that have recently migrated into the core, for which we find one example in the Pop. 1 stars of our simulation ($m_*\approx 0.3\, M_\odot$, $\dot{M}_\mathrm{wind}\approx 4\times 10^{-7}\, M_\odot$~yr$^{-1}$ in Figure~\ref{subfig:msmdot}). All other Pop. 1 PPDs either experience much lower UV flux, or have previously been depleted down to radii at which photoevaporation is not significant. 
 
  \begin{figure*}
 \centering
       \subfloat[\label{subfig:rIF} Proplyd ionisation front radius $R_\mathrm{IF}$]{\includegraphics[width=0.46\textwidth]{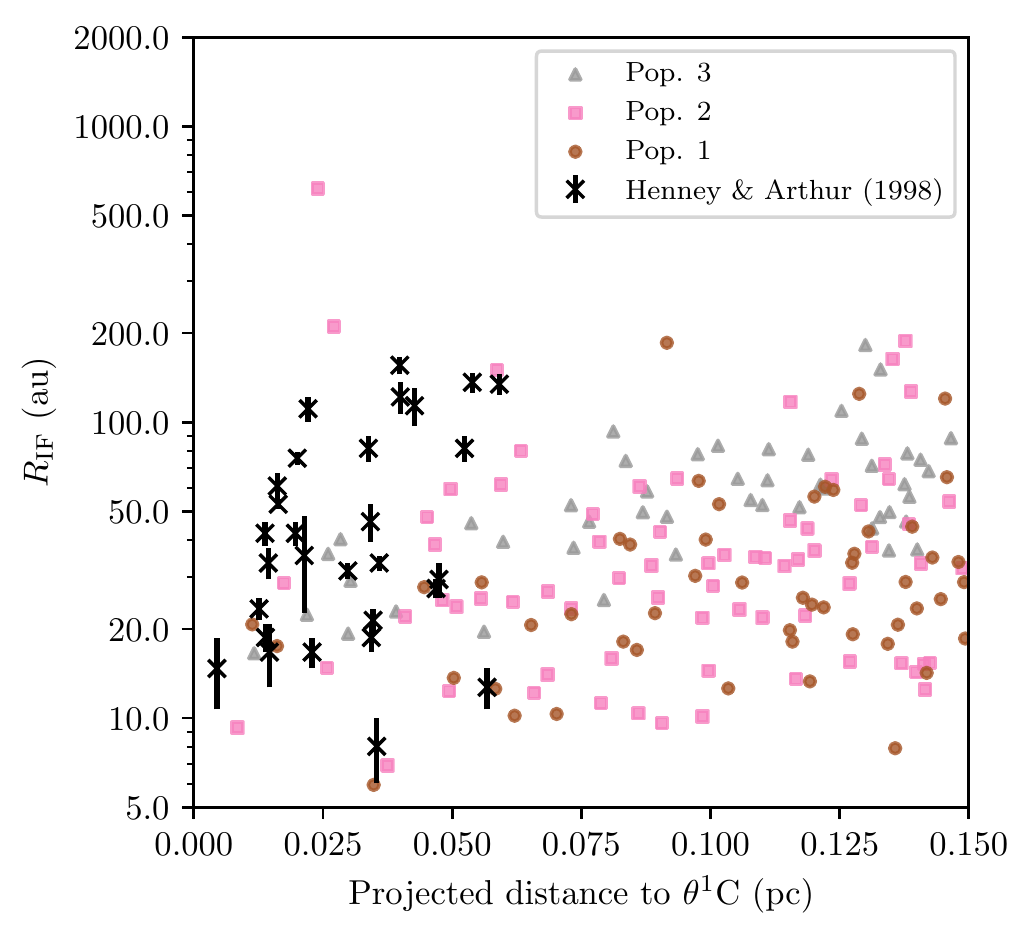}}
       \subfloat[\label{subfig:SHa} Mean proplyd H$\alpha$ surface brightness $\langle S(\mathrm{H}\alpha) \rangle$]{\includegraphics[width=0.45\textwidth]{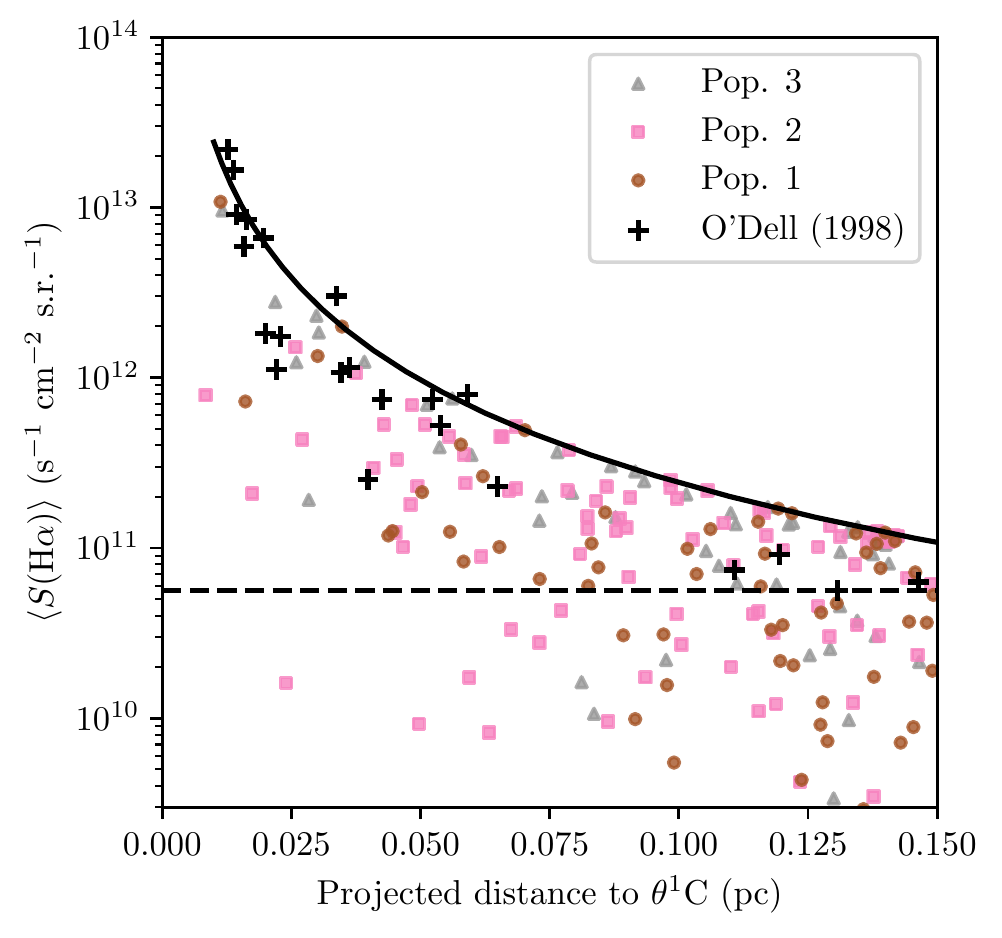}}
     \caption{In Figure~\ref{subfig:rIF} we show the radius of the IF against distance to $\theta^1$C (analogue). The simulation results are represented by scatter points in the same way as previous figures, with $R_\mathrm{IF}$ calculated using equation~\ref{eq:rIF}. \citet{Hen98} discuss that the sample is not complete for compact proplyds with $R_\mathrm{IF}\lesssim 50$~au. In Figure~\ref{subfig:SHa} we instead show the mean surface brightness $\langle S(\mathrm{H}\alpha)\rangle$, with the \citet{Ode98} sample shown as black crosses. The lowest surface brightness in that sample is represented by a horizontal dashed line. The solid black line follows the theoretical surface brightness assuming the projected separation from $\theta^1$C is the physical separation. }  
     \label{fig:rIF}
\end{figure*}

Although mass loss rates are not measured uniformly across the proplyds in the core of the ONC, and therefore a  full statistical comparison is not possible, we can make some quantitative comparisons between our results and the available samples. We compare our values for $\dot{M}_\mathrm{wind}$ with the findings of \mbox{\citet{Hen99}}, who obtained spectroscopic measurements of four proplyds (preferentially chosen for their brightness) to obtain a flow velocity and density. This analysis yielded mass loss rates in the range $ 7\times 10^{-7}$--$1.5 \times 10^{-6}\, M_\odot$~yr$^{-1}$, with a factor $2$ uncertainty. Assuming the same flow velocity for the larger sample of $27$ proplyds catalogued by \mbox{\citet{Hen98}}, the inferred mass loss rates are $\dot{M}_\mathrm{wind} \sim 10^{-8}$--$10^{-6}\, M_\odot$~yr$^{-1}$, with a mean of $\sim 3\times 10^{-7}\, M_\odot$~yr$^{-1}$. We obtain a lower $\dot{M}_\mathrm{wind}\sim 5\times 10^{-8}\, M_\odot$~yr$^{-1}$ for the majority of the Pop. 3 PPDs. However, this is consistent given the factor $2$ uncertainty and that the observed samples were preferentially chosen to be bright and extended (those proplyds with the greatest mass loss rates). Such mass loss rates may not be typical of the wider sample; indeed, the ionised flow from the proplyd LV 2 was later measured by \citet{Hen02}, which has a loss rate $\dot{M}_\mathrm{wind}\approx 10^{-8}\,M_\odot$~yr$^{-1}$. In terms of the number of PPDs that are rapidly losing mass, we emphasise that we do not obtain mass loss rates over the full IMF since we are limited in the stellar mass range of the \textsc{Fried} grid. We may therefore underestimate the number of PPDs which would exhibit mass loss rates $>10^{-7}\,M_\odot$~yr$^{-1}$.

Another important metric for comparison with the proplyd sample is the ionisation front (IF) radius $R_\mathrm{IF}$. This radius can be calculated from $\dot{M}_\mathrm{wind}$ and the EUV flux independently of whether the wind is driven by EUV or FUV photons \mbox{\citep[see discussion in][]{Cla15}}:
\begin{equation}
\label{eq:rIF}
    R_\mathrm{IF} = 21.5 \left(\frac{d}{0.1 \, \mathrm{pc}} \right)^{2/3} \left(\frac{\dot{M}_\mathrm{wind}}{10^{-8} \, M_\odot \, \mathrm{yr}^{-1}} \right)^{2/3} \left( \frac{\Phi_{\mathrm{i}}}{10^{49} \, \mathrm{s}^{-1}} \right)^{-1/3} \, \mathrm{au}
\end{equation} where $d$ is the distance to the ionising source (equation ~\ref{eq:rIF} is appropriate at the end of our simulation, where we neglect the influence of interstellar extinction). The results of this calculation are shown in Figure~\ref{subfig:rIF}, where we also show the IF radii calculated by \citet{Hen98}, who infer physical quantities using detailed models and the HST data collected by \mbox{\citet{Ode98}}. We do not, however, compare with the \citet{Vic05} sample, since the authors did not extract physical quantities from a model. Hence direct comparison would require synthetic images for each of the discs in our model. For the proplyds which are common to both samples, the values for $R_\mathrm{IF}$ are not consistent and it is therefore not possible to extrapolate between them. We find that, while there are a number of marginally more extended proplyds at small separations from $\theta^1$C in the \citet{Hen98} sample, we also find examples of large $R_\mathrm{IF}$ for small $d$ in our model. Drawing comparisons between the mean $R_\mathrm{IF}$ is not useful, since the observed sample is incomplete for $R_\mathrm{IF}\lesssim 50$~au and may also be limited by the surface brightness of the proplyds at large $d$. 

To explore the issue of proplyd surface brightness, we consider the original HST observations presented by \citet{Ode98}. The average surface brightness of $\mathrm{H}\alpha$ can be linked to the EUV flux:
\begin{equation}
    \label{eq:SHalpha}
    \langle S(\mathrm{H}\alpha) \rangle  \approx 7 \times 10^{11} \frac{\alpha_{\mathrm{H}\alpha}^{\mathrm{eff}}}{\alpha_B}\frac{\Phi_\mathrm{i}}{10^{49} \, \mathrm{s}^{-1}} \left( \frac{d}{0.1 \, \mathrm{pc}} \right)^{-2} \, \mathrm{s}^{-1} \, \mathrm{cm}^{-2} \, \mathrm{s.r}^{-1}
\end{equation} where $\alpha_{\mathrm{H}\alpha}^\mathrm{eff} = 1.17\times 10^{-13}$~cm$^{-3}$~s$^{-1}$ and $\alpha_B = 2.59\times 10^{-13}$~cm$^{-3}$~s$^{-1}$ at temperature $T=10^4$~K are hydrogen recombination coefficients \citep[][]{Ost89, Dra11}. The result of this calculation for our sample is shown in Figure~\ref{subfig:SHa} for proplyds as a function of separation to $\theta^1$C compared with the \citet{Ode98} sample. We again find comparable numbers of bright proplyds at small $d$, however the model predicts a larger number than found in observations at large $d$. In this case, some bright proplyds are also not extended such that the observational sample is resolution limited. Our results highlight that the available catalogues may be incomplete.

Taken together, the findings we have presented in this section illustrate another aspect of the proplyd lifetime problem: that the inferred mass loss rates for proplyds quoted in the literature are preferentially high. In reality, the most extreme values of $\dot{M}_\mathrm{wind}\sim 10^{-7}$--$10^{-6}\, M_\odot$~yr$^{-1}$ are probably only sustained for very short periods where the disc remains extended in a high UV flux environment. The majority of proplyds in our models only have a factor $2$--$3$ lower $\dot{M}_\mathrm{wind}$ than the extreme cases, and this lower loss rate persists for the vast majority of the disc lifetime. The difference is therefore important in understanding the dispersal timescale for the young discs which occupy the central regions of the ONC. 

\section{A solution to the proplyd lifetime problem}
\label{sec:PLP_sol}
\subsection{Summary of the solution}

In the previous section, we have presented many aspects of the properties of the ONC which contribute to understanding the properties of the PPDs and proplyds which have been inferred. Our findings indicate that `the solution' to the proplyd lifetime problem cannot be attributed to a single consideration or missing physics, but a combination of many different contributing (and related) factors. These can be summarised as follows:
\begin{enumerate}
    \item \textit{A population of stars younger than the mean age of the ONC.} These younger stars have a larger reservoir of circumstellar mass remaining, and therefore contribute to the PPD survival fraction. 
    \item \textit{Radial orbits during periods of star formation.} If the stars form in a subvirial state with respect to the gas potential, then stars will fall towards the centre early during their evolution and are accelerated by dynamical encounters. As the gas is depleted by further star formation and outflows, older stars migrate outwards and younger stars inwards. For sufficiently low SFE this can lead to preferentially young stars (with the greatest disc mass reservoir) occupying the core, where external photoevaporation acts most strongly on PPDs. 
    \item \textit{Interstellar extinction.} Extinction due to the persistence of primordial gas in the region has a moderate influence on the PPD survival fraction. It has the strongest influence on the oldest stars, which are protected during the early stages of the formation of the ONC by high gas densities. More detailed modelling, such as the stellar feedback calculations performed by \citet{Ahm19}, are required in the context of extended or discrete epochs of star formation. 
    \item \textit{Disc outer radius depletion.} A model of photoevaporating discs must take into account that the outer radius of the disc, from which the wind is launched, is not constant over time. Since external photoevaporation is more efficient at larger radii, where the gravitational potential is weaker, fixing the outer radius of the disc will lead to underestimating the disc destruction timescale. 
    \item \textit{Observational biases.} In particular, all studies of proplyds have preferentially targeted bright and extended examples, which also exhibit the highest mass loss rates ($\gtrsim 10^{-7}\, M_\odot$~yr$^{-1}$). While we find comparable numbers of IFs that are bright and extended, the vast majority of PPDs have much lower mass loss rates (a few $10^{-8}\,M_\odot$~yr$^{-1}$) which is likely typical of the wind driven mass loss rate averaged over a disc lifetime. The brightest proplyds are probably around stars with extended discs that have recently migrated into regions of strong EUV fields, and the mass loss rates inferred for them cannot be extrapolated to the rest of the population. This is similar to the radial orbit hypothesis put forward by \citet{Sto99}. In our models, however, only a small number of stars (whose discs exhibit the highest mass loss rates) must be on radial orbits; the majority of stars in the core survive because they are preferentially young \citep[point ii -- c.f.][]{Sca01}.
\end{enumerate}

\subsection{Caveats}

The modelling procedure we have presented makes a number of simplifications and is subject to caveats that we discuss in this section. Our intention in this work is to illustrate the mechanism by which PPDs in the ONC can survive; here we indicate how future work may consider these arguments in greater detail \citep[for example with hydrodynamic and radiative transfer modelling -- e.g.][]{Ahm19}.

\subsubsection{Kinematic model}
\label{sec:caveats_kin}
Our kinematic model is a simplified `toy model' for the dynamical evolution of the ONC, meant to capture the aspects that are important for our consideration of the PPD population. We have assumed a Plummer sphere gas potential that changes instantaneously during an `epoch' of star formation, and neglected any substructure in both the gas and the stellar population. Physically, we would expect star formation to continue over a finite period of time, and the quantity of gas in the region may have been constantly changing due to inflows and outflows. Our assumptions, including that star formation happens at discrete intervals, are simplifying but not necessary for the success of our model. The motivation here is finding a solution for which the youngest stars occupy the centre at the present time, which would be satisfied by any model that allowed stars to form significantly subvirially to the gas over an extended period. Since such a model would necessarily undergo cold collapse at early times, unravelling initial conditions from present day stellar kinematics becomes impossible \citep{Par14}. Therefore any deviation from our simple model is empirically unconstrained.

One additional factor which we have not considered in our model is the influence of binaries. Since the stellar density in the core is $n_0\gtrsim 2 \times 10^4$~pc$^{-3}$ throughout the dynamical evolution of our model, a large number of multiple star encounters would be expected if a significant number of stars formed in binaries. Such scatterings could lead to a number of stars undergoing strongly radial orbits, which may add to the frequency of bright and extended proplyds at the present day (see Section~\ref{sec:caveats_PPDs}). The reason for not considering binaries is to minimize the stochasticity of our models so that comparisons could be drawn between parameter choices. We would not, for example, want our $\theta^1$C analogue to be scattered out of the core of the region for a significant length of time \citep{Pfl06}, since comparisons of the PPD populations between models would be problematic. We leave understanding the influence of binaries on the proplyd population for future work.

\subsubsection{PPD population}
\label{sec:caveats_PPDs}

As in the case of the kinematic model, the viscous disc evolution model is necessarily simplified. Having demonstrated that the discs which experience high mass loss rates are the minority (stars with extended discs which migrate towards the core), we consider what factors might alter the number of extended and bright proplyds at small separations from $\theta^1$C. Some such considerations are as follows:
\begin{itemize}
    \item The greatest limitation of our model is that we are unable to calculate FUV induced mass loss rates across the entire IMF. The \textsc{Fried} grid does not include mass loss rates for host masses $<0.3 \, M_\odot$, and this would increase the number of discs in our sample by a factor $\sim 2$. 
    \item  Initial disc masses and radii are fixed in our models. If some subset of discs are initially more massive or extended, then they could exhibit greater mass loss rates at the present day and may contribute to the number of bright and extended proplyds. 
    \item A clumpy gas distribution might result in greater protection for some PPDs. This is especially the case for young stars for which we have not considered the influence of extinction, but in reality must have formed in some overdensity of gas. In general, a clumpy gas distribution is less efficient at protecting a disc population as a whole, but may lead to individual PPDs which evolve without being strongly irradiated for an extended period.
    \item Our model does not include binaries, which may have a significant role in the dynamical evolution of the ONC (Section~\ref{sec:caveats_kin}). If stars are excited into radial orbits by scattering, then this would increase the number of stars which occupy the central regions of the ONC briefly, and therefore the number of bright proplyds with high mass loss rates at a given time. 
\end{itemize} While our simplified model does not factor in these considerations, we still obtain multiple examples of rapid mass loss rates ($\gtrsim 10^{-7}\, M_\odot$~yr$^{-1}$) and the corresponding bright and extended proplyds. Current observed samples are incomplete for $R_\mathrm{IF} \lesssim 50$~au and are subject to variable background noise due to the ionised gas anterior to the ONC \citep{Bal00}. Future samples of proplyds may be more complete such that the mass loss rate distribution can be better constrained.   
 
 In addition to the proplyd properties, a number of physical and observational considerations may influence the measured properties of the broader PPD population. Some of these considerations (e.g. uncertainties in stellar ages and early ram pressure stripping) are discussed in Section~\ref{sec:PPDsurv} in the context of disc survival. Further, the standard assumptions that apply to observed disc properties also apply to our model. When comparing to empirically derived disc masses, uncertainties in the opacity and temperature of the dust will alter the inferred distribution. Finally, the outer radius of the dust component of the disc does not necessarily trace the outer radius in the gas \citep{Ros19}. This is particularly the case if the radial drift timescale is shorter than timescale on which the disc is depleted by external photoevaporation. Future modelling of the dust properties in photoevaporating discs is required to understand the relationship between the dust and gas properties \citep[see][]{Sel19}.

 \section{Conclusions}
 \label{sec:concs}
 
 We have presented a simple dynamical model that reproduces the observed kinematics and distribution of stellar ages in the ONC, while also explaining the survival of the strongly irradiated PPD population in the central regions (the `proplyd lifetime problem'). The influence of a gas potential on the stellar populations that are born at different times yields a gradient of stellar ages (which are observed to be younger in the core). This has a two-fold influence on the external photoevaporation of discs. Firstly, the interstellar UV extinction can shield older PPD populations in their early evolution until the gas is expelled and they migrate away from $\theta^1$C. Secondly, the stars which are presently strongly irradiated are also the youngest, and therefore have the greatest mass reservoir remaining. This represents a natural mechanism by which discs in the ONC can survive until the present day, without needing to invoke the extreme youth of $\theta^1$C or initially extreme disc masses. In addition, we highlight that the mass loss rates quoted for observed proplyds are preferentially high, and that for the vast majority of its lifetime a disc in the ONC experiences lower wind driven mass loss rates of a few $10^{-8}\, M_\odot$~yr$^{-1}$. 

Our model also represents a test of the multiple stellar population hypothesis for the ONC. In the (projected) core, the differences between the disc property distributions across stellar populations of different ages are as follows:
\begin{enumerate}
\item The older population has on average less massive PPDs.
\item The relationship between disc mass and host mass steepens with age. While this is possibly also a signature of secular disc evolution, external photoevaporation is expected to enhance it. However, this finding has the major caveat that the relationship begins to flatten again when discs fall below the sensitivity limit and drop out of the sample. This leaves preferentially higher mass discs and creates a bias towards stars which are at a physically greater distance from the core than is apparent from their projection.
\item Linking to the previous point, PPDs with a significant dust mass in an older stellar population may be preferentially hosted by more massive stars with respect to the younger populations. 
\item Disc radii in each individual population exhibit stronger correlation with stellar host mass than is found across all populations.
\item PPDs in older populations have preferentially smaller discs than those in younger populations.
\end{enumerate}
Our results therefore represent both a solution to the proplyd lifetime problem and a test for the hypothesis that multiple stellar populations exist in the ONC. We conclude that because of the complex star formation history, and because it hosts a population of stars which are young ($\sim 0.8$~Myr old) with respect to the age spread, using the ONC as the archetypal region of study for externally depleted PPDs yields confusing and conflicting results. Regions such as Cygnus OB2, for which the spread of ages is smaller with respect to the average age of the region, show clearer signs of externally photoevaporated discs \citep{Gua16, Win19}.

\section*{Acknowledgements}

We thank the anonymous referee for a positive report. AJW thanks Richard Booth for useful discussion and suggestions. This project has received funding from the European Research Council (ERC) under the European Union's Horizon 2020 research and innovation programme (grant agreement No 681601). This project has received funding from the European Union's Horizon 2020 research and innovation programme under the Marie Sk\l{}odowska-Curie grant agreement No 823823 (DUSTBUSTERS). GR acknowledges support from the Netherlands Organisation for Scientific Research (NWO, program number 016.Veni.192.233). This work is part of the research programme VENI with project number 639.041.644, which is (partly) financed by the Netherlands Organisation for Scientific Research (NWO). AH thanks the Spanish MINECO for support under grant AYA2016-79006-P.




\bibliographystyle{mnras}
\bibliography{truncation} 


\appendix
\section{Host mass dependent initial disc mass}
\label{app:hostmass}

  \begin{figure}
 \centering
       \includegraphics[width=0.45\textwidth]{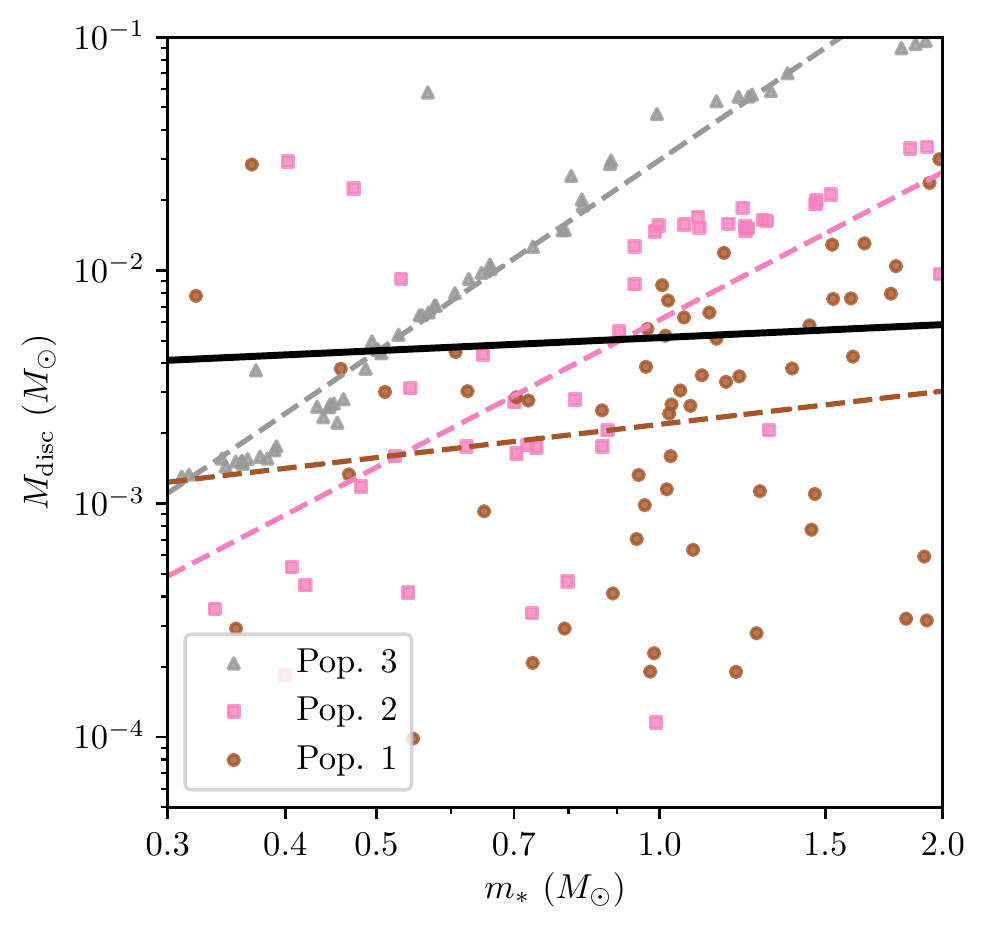}
     \caption{As in Figure~\ref{subfig:msmd_1e-3} but for initial conditions where the initial disc mass $M_{\mathrm{disc},0} = 0.1 m_*$ (as opposed to $M_{\mathrm{disc},0} = 0.1 \, M_\odot$). The power law index is $\beta=0.19$ for all stars and $\beta = 0.5$, $2.1$ and $2.7$ for Pops. 1, 2 and 3 respectively. }
     \label{fig:mdmhost_altics}
\end{figure}

During the course of this study we have assumed host mass independent disc initial conditions to investigate the influence of disc dispersal on the relationship $M_\mathrm{disc}(m_*)$. However, observations indicate that $M_\mathrm{disc}$ is approximately linearly (or moderately super-linearly) related to $m_*$ \citep[e.g.][]{Andr13, Pas16}. In this appendix, we consider our findings in Section~\ref{sec:MdiscvMstar} that disc and host masses appear uncorrelated when population membership is not accounted for. We ask whether this is still true for host mass dependent disc initial conditions. For this purpose, we run the same simulations but this time with $M_{\mathrm{disc},0} = 0.1 m_*$.

Figure~\ref{fig:mdmhost_altics} shows the result of this exercise in terms of the host mass--disc mass relationship after $2.8$~Myr. We again find a shallow relationship accross all populations, but a steeper (and in this case super-linear) relationship within the individual populations. Our finding that a correlation can be disguised by contamination by multiple stellar populations therefore holds in the case where $M_{\mathrm{disc},0} = 0.1 m_*$.

\bsp	
\label{lastpage}
\end{document}